%% file: AXEL2016.tex
\PassOptionsToPackage{pdftex}{graphicx} 
\documentclass[3p,twocolumn]{elsarticle}  
\usepackage{lineno,hyperref}
\usepackage[pdftex]{graphicx}
\modulolinenumbers[100]

\journal{Nuclear Instrument and Method A}

\begin{document} 
\begin{frontmatter}

\title{Electroluminescence collection cell as a readout for a high energy resolution Xenon gas TPC}

\author[kyoto]{S.Ban} \ead{bansei0526@scphys.kyoto-u.ac.jp}
\author[kyoto]{K.D.Nakamura} \ead{kiseki@scphys.kyoto-u.ac.jp}
\author[kyoto]{M.Hirose}
\author[kyoto]{A.K.Ichikawa}
\author[kyoto]{A Minamino}
\author[kobe]{K.Miuchi}
\author[kyoto]{T.Nakaya}
\author[kamioka]{H.Sekiya} 
\author[kyoto]{S.Tanaka}
\author[tohoku]{K.Ueshima}
\author[kyoto]{S.Yanagita}
\author[kyoto]{Y.Ishiyama}
\author[kyoto]{S.Akiyama}
\address[kyoto]{Kyoto University, Kitashirakawaoiwake-cho Sakyo-ku Kyoto-shi Kyoto, 606-8502, Japan}
\address[kamioka]{Kamioka Observatory, ICRR, The University of Tokyo, 456 Higashimozumi Kamioka-cho Hida-shi Gifu, 506-1205, Japan}
\address[tohoku]{RCNS, Tohoku University, 6-3 Aramakiazaaoba, Aoba-ku Sendai-shi, Miyagi, 980-8578, Japan}
\address[kobe]{Kobe University, Rokodai, Nada-ku Kobe-shi, Hyogo, 657-8501, Japan}

\begin{abstract}
AXEL is a high pressure xenon gas TPC detector being developed for neutrinoless double-beta
 decay search.
We use the proportional scintillation mode with a new electroluminescence light detection
 system to achieve high energy resolution in a large detector.
The detector also has tracking capabilities, which enable significant background rejection.
To demonstrate our detection technique, we constructed a $10\,\rm{L}$ prototype detector filled with up to
 $10\,\rm{bar}$ xenon gas.
The FWHM energy resolution obtained by the prototype detector with 4 bar Xenon gas is 4.0$\pm$0.30 \% at $122\,\rm{keV}$,
which corresponds to $0.9\,\sim\,2.0\%$ when extrapolated to the Q value of the $0\nu\beta\beta$
 decay of $^{136}$Xe.
\end{abstract}

\begin{keyword}
neutrinoless double beta decay, xenon, electroluminescence, time projection chambers
\end{keyword}

\end{frontmatter}

\linenumbers

\input{AXEL2016_Introduction}

\input{AXEL2016_ELCC}

\input{AXEL2016_Prototype}

\input{AXEL2016_Analysis}

\input{AXEL2016_Results}

\section{Conclusion}
AXEL is a high pressure xenon gas TPC designed to search for $0\nu\beta\beta$. It is the first detector to employ a cellular structure to collect ionizing electrons and detect electroluminescence light.
The geometry and electric field of the cell structure, ELCC, was optimized via simulations and the performance was demonstrated with a prototype detector. It was confirmed by simulation that the electron collection efficiency along the field lines is 100\% when proper electric field is applied.
The effect of the EL yield fluctuation on energy resolution is estimated to be less than $0.5\%$, and is therefore sufficiently small.
With the prototype detector, an FWHM energy resolution of 4.0$\pm$0.30 \% is achieved at
 122\, \rm{keV}, which corresponds to $0.9\,\sim\,2.0\%$ when extrapolated to the Q value
 of $0\nu\beta\beta$ decay of $^{136}$Xe.

\section*{Acknowledgments}
We thank R. Wendell for his support to prepare this paper.
This work was partially supported by
JPSP KAKENHI Grant Number JP15H02088. 

\bibliographystyle{elsarticle-num}
\bibliography{AXEL2016}

\end{document}

%% file: AXEL2016_Introduction.tex
\section{Introduction}
Observation of neutrinoless double beta decay ($0\nu\beta\beta$) is important to reveal the
 nature of the neutrino, such as the neutrino mass hierarchy, its absolute mass and 
 whether or not it is a Majorana particle\cite{ref:DBD_theory_1982}.
Among potential double beta decay nuclei, $^{136}\rm{Xe}$ offers several advantages in terms
 of detecting this process.
The natural abundance of $^{136}\rm{Xe}$ is as high as $8.9\%$ and can be enriched using
 established methods.
Very high energy resolution is possible in gaseous xenon, in principle, due to its large
 ionization yield and small fano-factor.
It also emits scintillation light.
The EXO experiment uses xenon and obtained $1.1\times 10^{25}\,\rm{yrs}$ as the
 $90\%\,\rm{C.L.}$ lower limit of the $0\nu\beta\beta$ half life\cite{ref:EXO_NATURE2014}.
The KamLAND-Zen experiment obtained $1.07\times 10^{26}\,\rm{yrs}$ as the $90\%\,\rm{C.L.}$
 lower limit using xenon dissolved in liquid scintillator\cite{ref:KamLAND_PRL2016}.
Longer half-life corresponds to lighter neutrino mass, and to further explore smaller neutrino
 mass up to so-called inverted mass ordering, sensitivity has to reach
 $6\times 10^{27}\,\rm{yrs}$ and energy resolution improvement is essential for discriminating
 radioactive and $2\nu\beta\beta$ backgrounds.
A $0\nu\beta\beta$ search using high pressure gaseous xenon was proposed\cite{ref:Chinowsky2007} in order to obtain high energy resolution and topological information and experiments have started (NEXT\cite{ref:NEXT_JINST2012}) or planned (PandaX-III\cite{ref:PANDAXIII_arXiv2016}).
The former experiment utilizes the electroluminescence and the latter the micro pattern gas detector to amplify the ionization electron signal. In these projects, high energy resolutions, 1.82\% (FWHM) at 511 keV\cite{ref:NEXT_511keV} and 9.6\% (FWHM) at 22.1 keV\cite{ref:micromegas}, have been demonstrated.
Our detector, a high pressure xenon gas TPC, AXEL, adopts a new method to measure ionization electrons by electroluminescence with a cellular structure, which enables high energy resolution for large target mass while maintaining strong background rejection power.
\par
\begin{figure}[htbp]
\centering
\includegraphics[width=.95\linewidth]{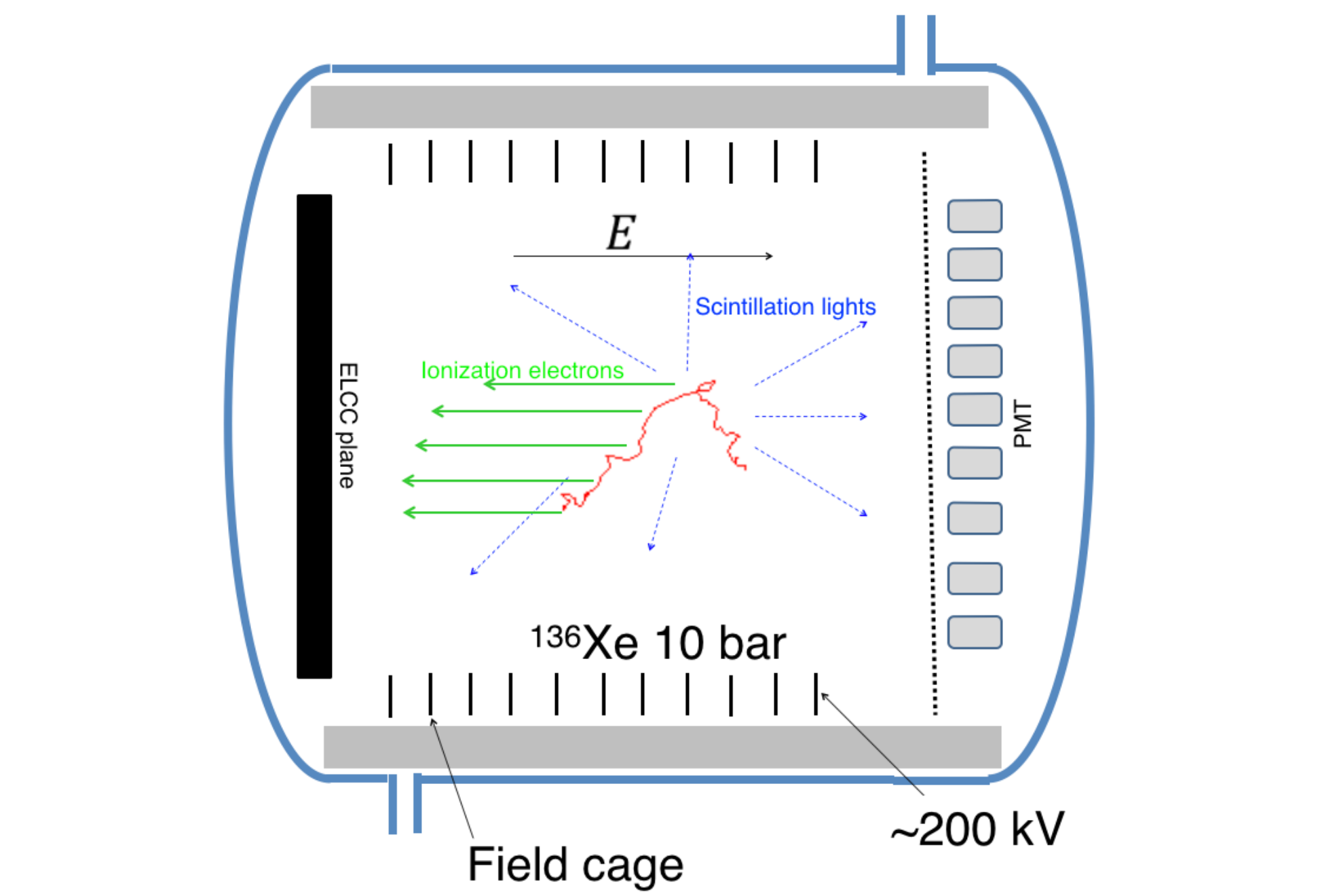}
\caption{ Schematic drawing of AXEL detector. }
\label{fig:AXEL}
\end{figure}
The schematic view of the AXEL detector is shown in Fig. \ref{fig:AXEL}.
It is a high pressure xenon gas TPC filled with $10\,\rm{bar}$ $^{136}\rm{Xe}$ enriched gas.
Ionized electrons are detected by a pixelized readout plane named ELCC (ElectroLuminescence
 Collection Cell, described in Sec. \ref{sec:ELCC}) placed at electron drifting side.
Scintillation light is detected by PMTs on the opposite side of the vessel to obtain the hit
 timing which is necessary for event fiducialization.
In the past, $0.3\%$ (FWHM) energy resolution for the $662\,\rm {keV}$ gamma ray was
 demonstrated\cite{ref:Bolotnikov_NIM1997} for ionization chamber filled with xenon gas.
We aim for $0.5\%$ as a realistic energy resolution with large volume by adopting the ELCC
 readout.
In this paper, we describe the concept of the ELCC and report its first performance result.

%% file: AXEL2016_ELCC.tex
\section{Electroluminescence Light Collection Cell (ELCC)}
\label{sec:ELCC}

\subsection{Concept}
\begin{figure}[t]
\centering
\includegraphics[width=.8\linewidth]{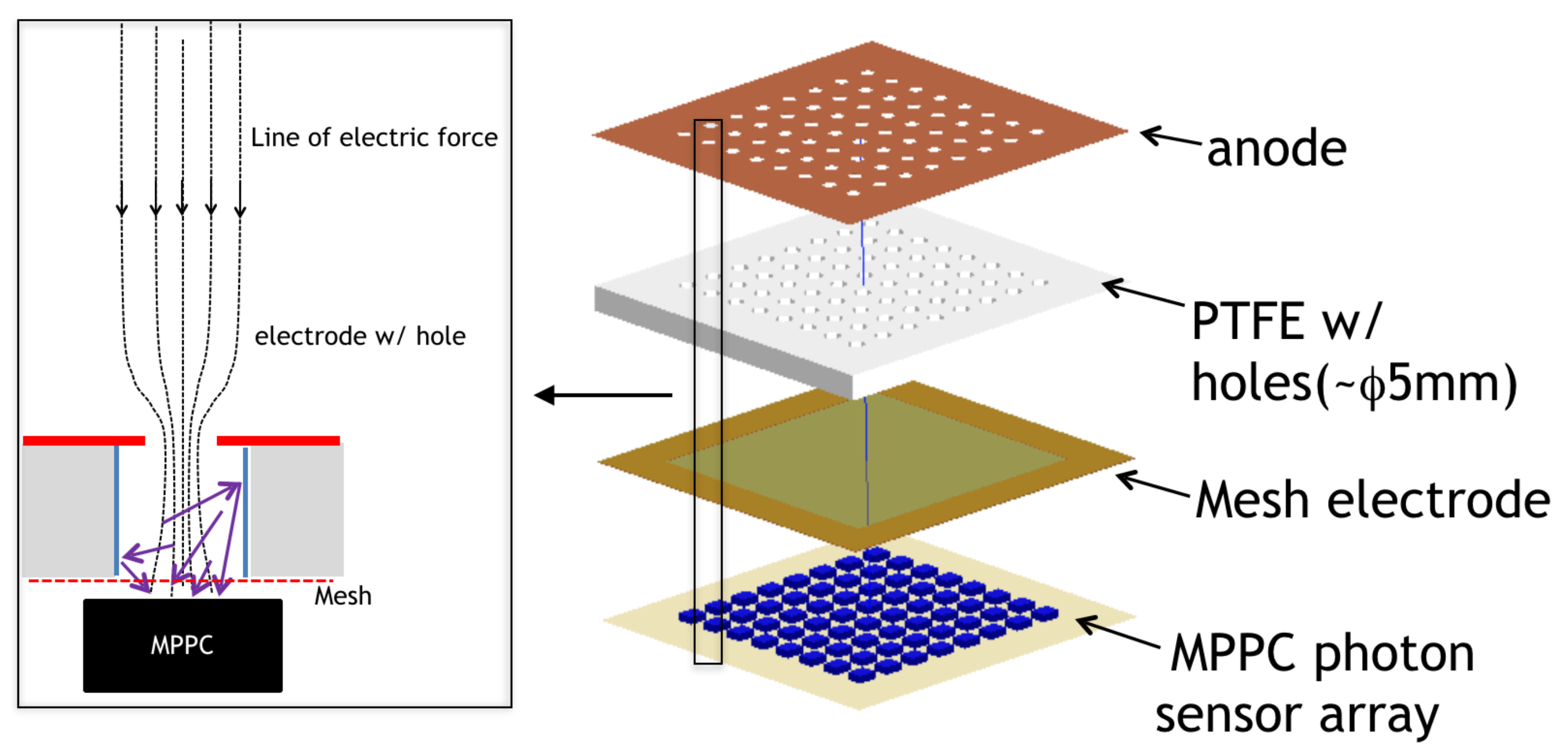}
\caption{ Structure of ELCC. }
\label{fig:ELCC}
\end{figure}
Electroluminescence (EL) is a process in which electrons accelerated in a high electric field
 excite xenon atoms and generate de-excitation photons.
The EL photons are always generated by the initial electron unlike the avalanche amplification
 where initial fluctuation is amplified, too.
A normal way to utilize the EL process for the radiation detection is applying high voltage
 between two conductive parallel meshes to generate EL photons.
Those photons are detected by photon sensors such as PMT.
Such systems exhibit good energy resolution in compact detectors\cite{ref:BeppoSax_AA1997}.
However, when the detector volume is large, it is difficult to get uniform coverage by photon
 sensors and the energy resolution is worsened because the acceptance to detected photon
 depends on the position of radiation inside the detector volume.
To solve this problem, we propose the ELCC.
\par
ELCC is designed to measure both energy deposition and event topology.
Figure \ref{fig:ELCC} depicts the structure of ELCC.
The EL region is made of a Cu plate, PTFE plate and mesh.
The PTFE plate and Cu plate has holes to form cells.
For each cell, a SiPM photo-sensor is attached at the back of the mesh electrode to detect EL
 photons.
The mesh is electrically connected to ground and negative voltage ($\sim -15\,\rm{kV}$)
 is applied to the Cu plate.
The space above the ELCC is the target volume, whose drift field uses the Cu plate of the ELCC
 as its anode.
By applying sufficiently high voltage between the anode electrode and mesh, ionized
 electrons are collected into cells along the lines of electric field, and generate
 EL photons, which are detected by SiPMs in each cell.
Because the acceptance of the SiPM for the EL light does not depend on the event position in
 the TPC, ELCC measure number of ionized electrons without any event-position correction.
Also, since ELCC is pixelized, it would enable strong background rejection by the event
 topology.
Furthermore, the detection surface can easily be extended to larger areas due to the solid
 structure of the ELCC.
In this paper, $z$ axis is the direction of electron drift, and the $x$ and $y$ axes are
 parallel to the ELCC plane.
In this section, the origin of coordinates is intersection of the central axis of the cell in
 $x$-$y$ plane and anode-Cu plane in $z$ axis.
\par
\begin{figure}[t]
\centering
\includegraphics[width=.8\linewidth]{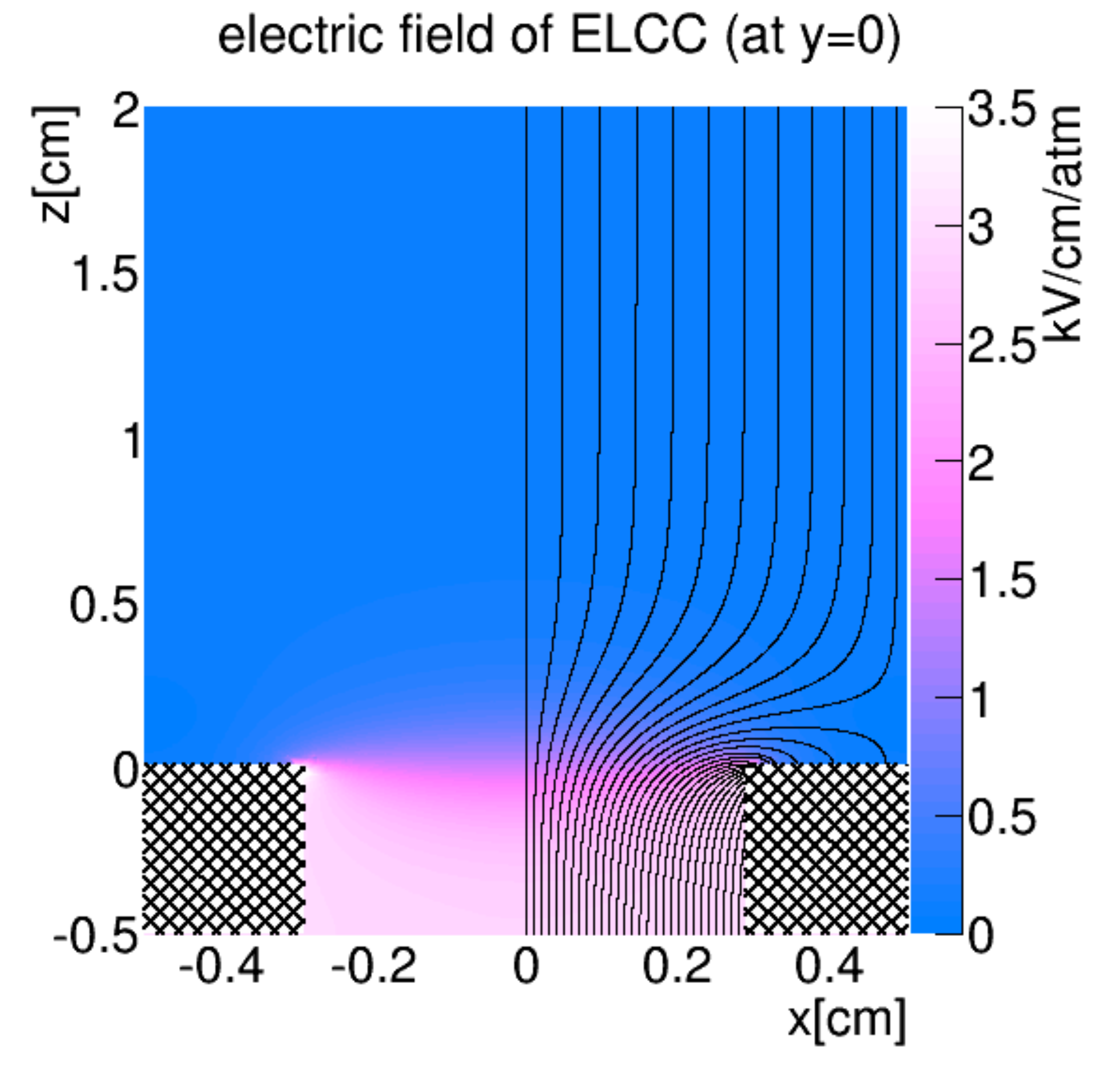}
\caption{ Calculated electric field at $y=0$ plane when voltage is applied at
 $100\,\rm V/cm/atm$ in the drift region and $3\,\rm kV/cm/atm$ in the EL region.
 Since the basic performance of the gas detector depends on the electric field normalized by pressure, the electric field strength as a color bar is expressed in units of V/cm/atm.
The horizontal and vertical axis correspond to $x$ and $z$ axis of the ELCC.
Colored contours show the reduced electric field strength.
The hatched regions correspond to the PTFE insulator with a $r=3\,\rm{mm}$ hole.
Electric field lines have additionally been drawn on the right half of the figure. }
\label{fig:field}
\end{figure}

\subsection{Optimization}
To determine the optimum voltage and geometrical parameters, we simulated the electric field
 with the finite element method (Gmsh\cite{ref:gmsh} and Elmer\cite{ref:elmer}).
The baseline geometry has a $10\,\rm mm$ cell pitch with a $5\,\rm mm$ deep EL region and a
 $6\,\rm mm$ diameter hole.
The cell pitch will be optimized from the track reconstruction ability and total cost of
 SiPMs and readout electronics.
Since ionization electrons diffuse about $10\,\rm{mm}$ for $1\,\rm{m}$ of drift in xenon gas,
 a $10\,\rm{mm}$ cell pitch is sufficiently fine, and anything smaller than that is not
 necessary.
In order to maintain the mechanical strength of PTFE insulator, at least $5\,\rm{mm}$ is
 required for the EL region.
It will be confirmed in Section \ref{sec:performance_estimation} that it is possible to obtain
 the sufficient number of EL photons for this length.
\par
\begin{figure}[t]
\centering
\includegraphics[width=.9\linewidth]{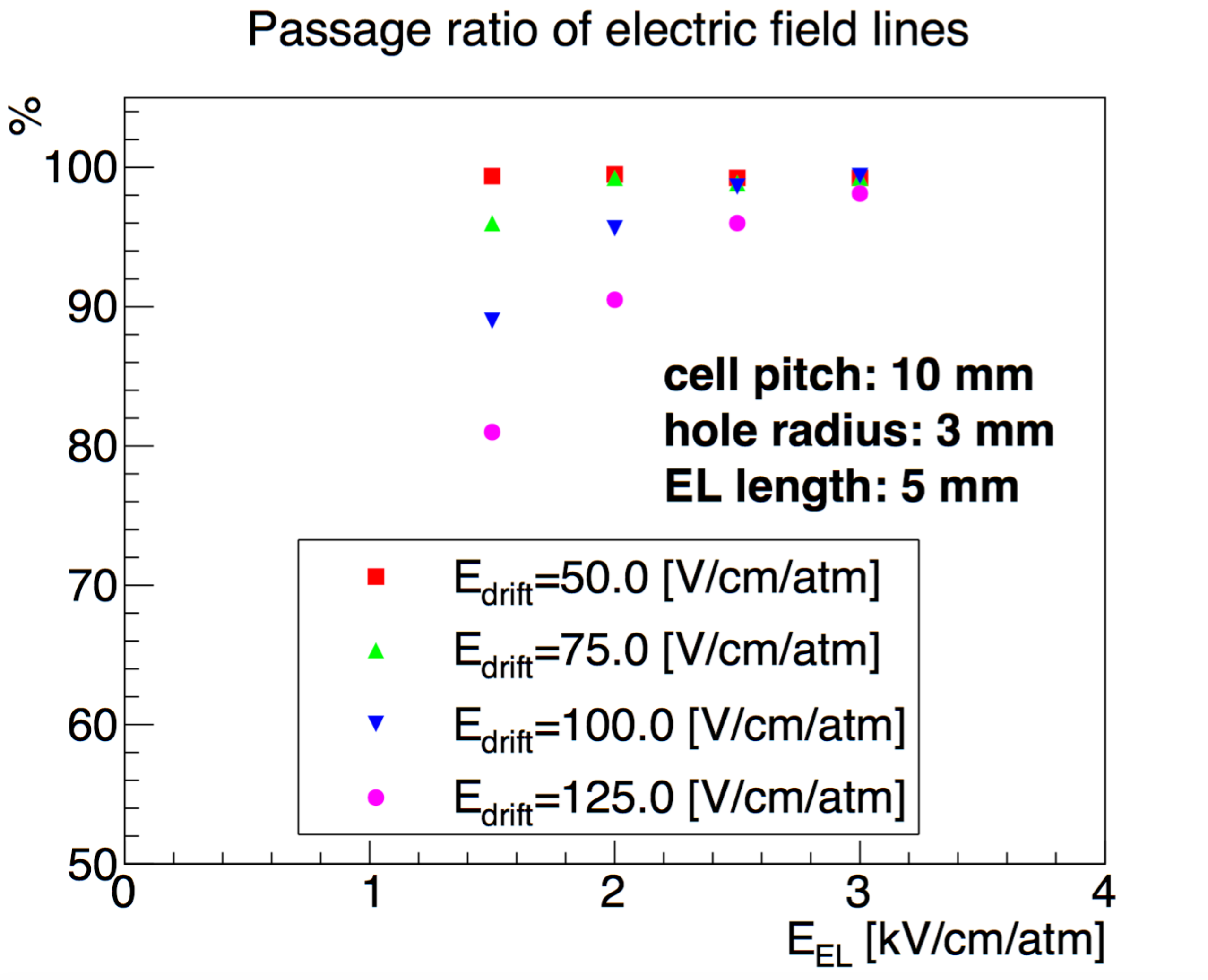}
\caption{ Electric field ($E_{\rm drift}, E_{\rm EL}$) dependence of the collection
 efficiency of line of electric field. }
\label{fig:pass_ratio_for_field}
\end{figure}
\begin{figure}[t]
\centering
\includegraphics[width=.95\linewidth]{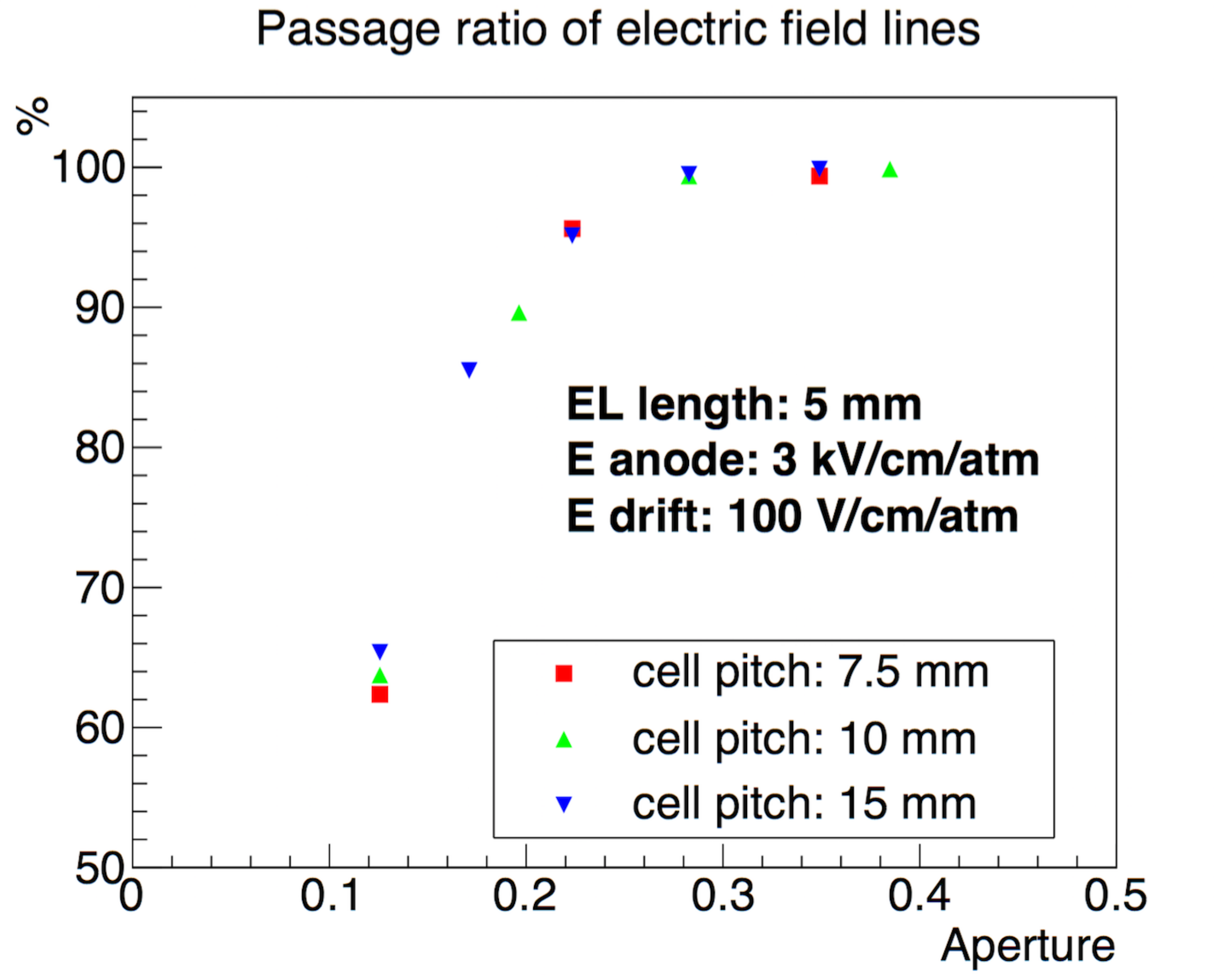}
\caption{ The cell geometry (cell pitch, hole radius) dependence of the collection
 efficiency of line of electric field. }
\label{fig:pass_ratio_for_geometory}
\end{figure}

Figure \ref{fig:field} shows an example of the calculated electric field distribution.
All electric field lines converge on the ELCC hole.
Figure \ref{fig:pass_ratio_for_field} shows the electric field dependence of the collection
 efficiency of electric field lines defined as the percentage of electric field lines
 generated above $2\,\rm cm$ of ELCC going into the hole.
The efficiency is better for the stronger EL field and weaker drift field.
To suppress recombination and to get good energy resolution, the drift field higher than
 $100\,\rm{V/cm/atm}$ is desired.
Thus, in order to maintain $100\%$ collection efficiency an EL field of
 $2.5\sim 3\,\rm{kV/cm/atm}$ is required.
Figure \ref{fig:pass_ratio_for_geometory} shows the dependence of the collection efficiency on
 cell geometry.
The efficiency depends on the aperture ratio of the ELCC hole defined as
 $(\pi r_{\rm{hole}}^2)/l_{\rm{pitch}}^2$, where $l_{\rm{pitch}}$ is the cell pitch and
 $r_{\rm{hole}}$ is the hole radius.
Based on the figure, the aperture ratio is required to be larger than $\sim 0.3$ to obtain
 $100\%$ collection efficiency.
When adopting $10\,\rm{mm}$ of cell pitch, the hole size should be more than $6\,\rm{mm}$ in
 diameter.
However, a $7.5\,\rm{mm}$ cell pitch and a $4\,\rm{mm}$ hole diameter was adopted for the
 prototype detector because diffusion is small with its shorter ($9\,\rm{cm}$) drift length as
 described in Section \ref{sec:prototype}.
\par

\begin{table}[bht]
\caption{Optimized ELCC parameters.} 
\centering
\begin{tabular}{c|c} \hline
  Parameter & Value \\ \hline
  Cell pitch & $10\,\rm{mm}$ \\
  EL region thickness & $5\,\rm{mm}$ \\
  Hole diameter & $6\,\rm{mm}$ \\
  EL region field & $3\,\rm{kV/cm/atm}$ \\
  Drift region field & $100\,\rm{V/cm/atm}$ \\
  \hline
\end{tabular}
\label{table:elcc_param}
\end{table}
\subsection{Performance estimation} \label{sec:performance_estimation}
\begin{figure}[t]
\centering
\includegraphics[width=.9\linewidth]{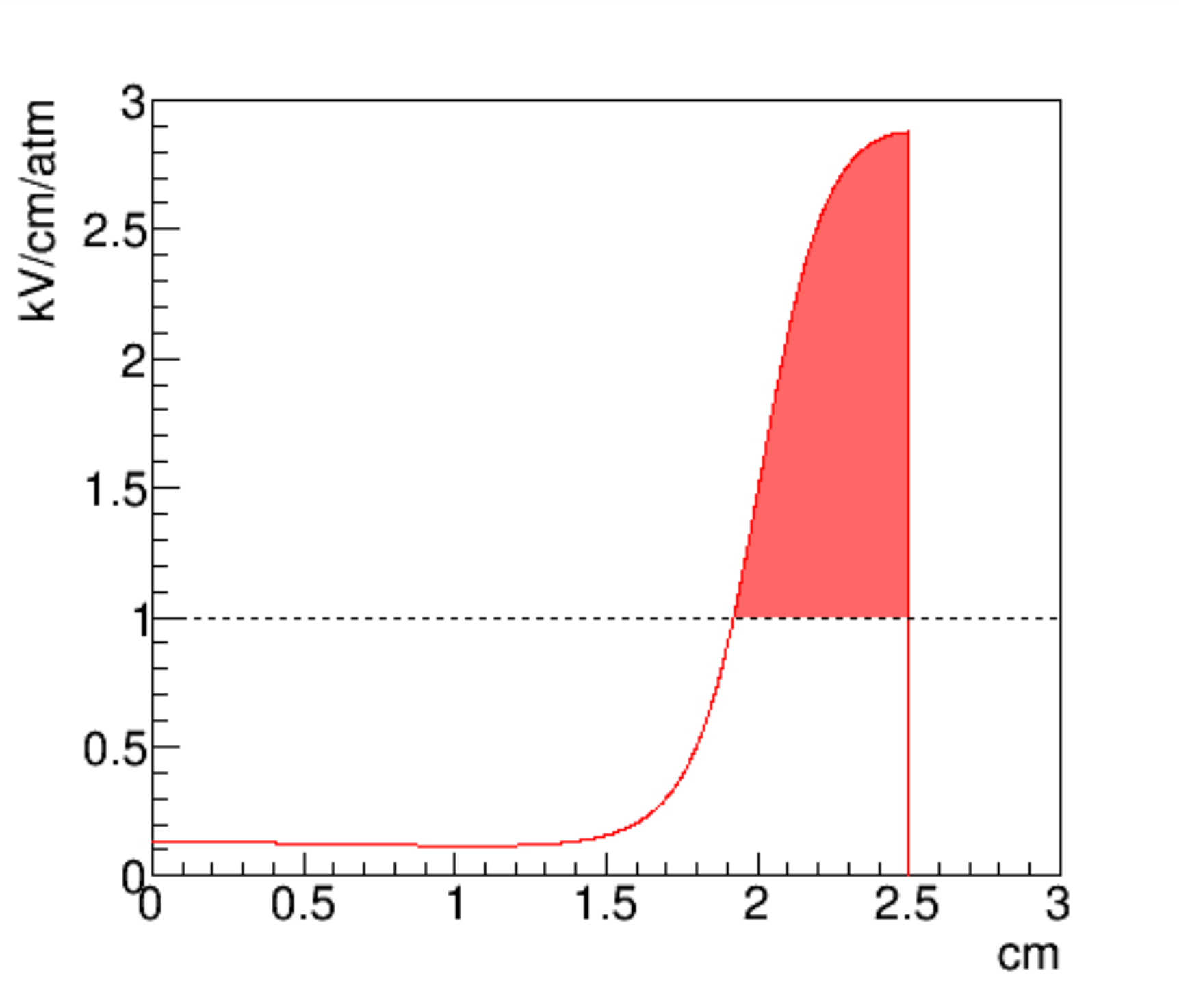} 
\caption{ Electric field strength along one of the lines of electric field.
The electroluminescence yield would be proportional to the hatched area. }
\label{fig:fieldAlongLine}
\end{figure}

\begin{figure}[t]
\includegraphics[width=.95\linewidth]{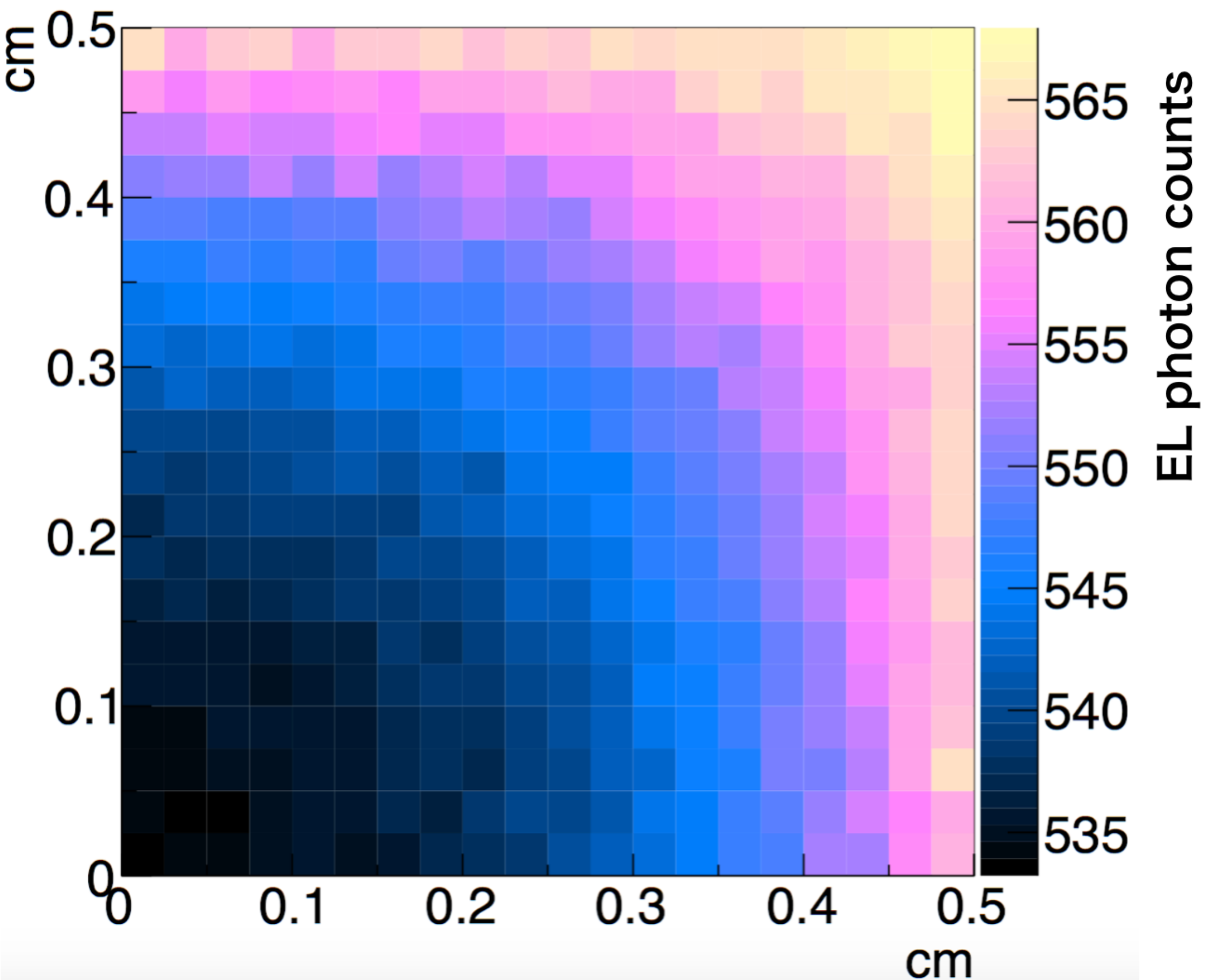} 
\caption{ Integral of the electrical field above the EL threshold along the field line.
The $x$ and $y$ axes shows the initial position at $2\,\rm{cm}$ above the anode plate. }
\label{fig:fieldIntegralMap}
\end{figure}

For the optimized geometry and electric field shown in Table \ref{table:elcc_param}, we
 investigated the uniformity of the EL light yield in a cell.
This uniformity is important to preserve good energy resolution.
Figure \ref{fig:fieldAlongLine} shows the electric field strength along one of the lines of
 electric field.
The average number of EL photons produced by one drifting electron is described by the following
 formula\cite{ref:NobleGasDetector}
\begin{equation}
dN_{\rm EL}/dx=70(E/p-1.0)p,
\end{equation}
where $x\,\rm[cm]$ is the path length of the electron, $E/p\,\rm [kV/cm/bar]$ is the reduced
 electric field strength, and $p\,\rm [bar]$ is the gas pressure.
EL photons are produced when electric field is stronger than the EL
 generating threshold ($1\,\rm kV/cm/bar$), and the number of photons is proportional to the
 electric field strength above the threshold.
To calculate the uniformity, for each of 400 initial positions at $2\,\rm{cm}$ above the Cu
 anode plate, the expected
 number of EL photons is calculated by integrating the electric field above the EL threshold
 along the electric field line (see Figure \ref{fig:fieldAlongLine}).
The calculated integral of the field is shown in Figure \ref{fig:fieldIntegralMap} as a function
 of the initial position.
Though the field integral shows concentric distribution, variation is as small as
 $1.7\%$ (rms).
The contribution of this non-uniformity to the energy resolution is expected to be very small and actually calculated to contribute 0.0054 \% by simulation.


\begin{figure}[t]
\centering
\includegraphics[width=.9\linewidth]{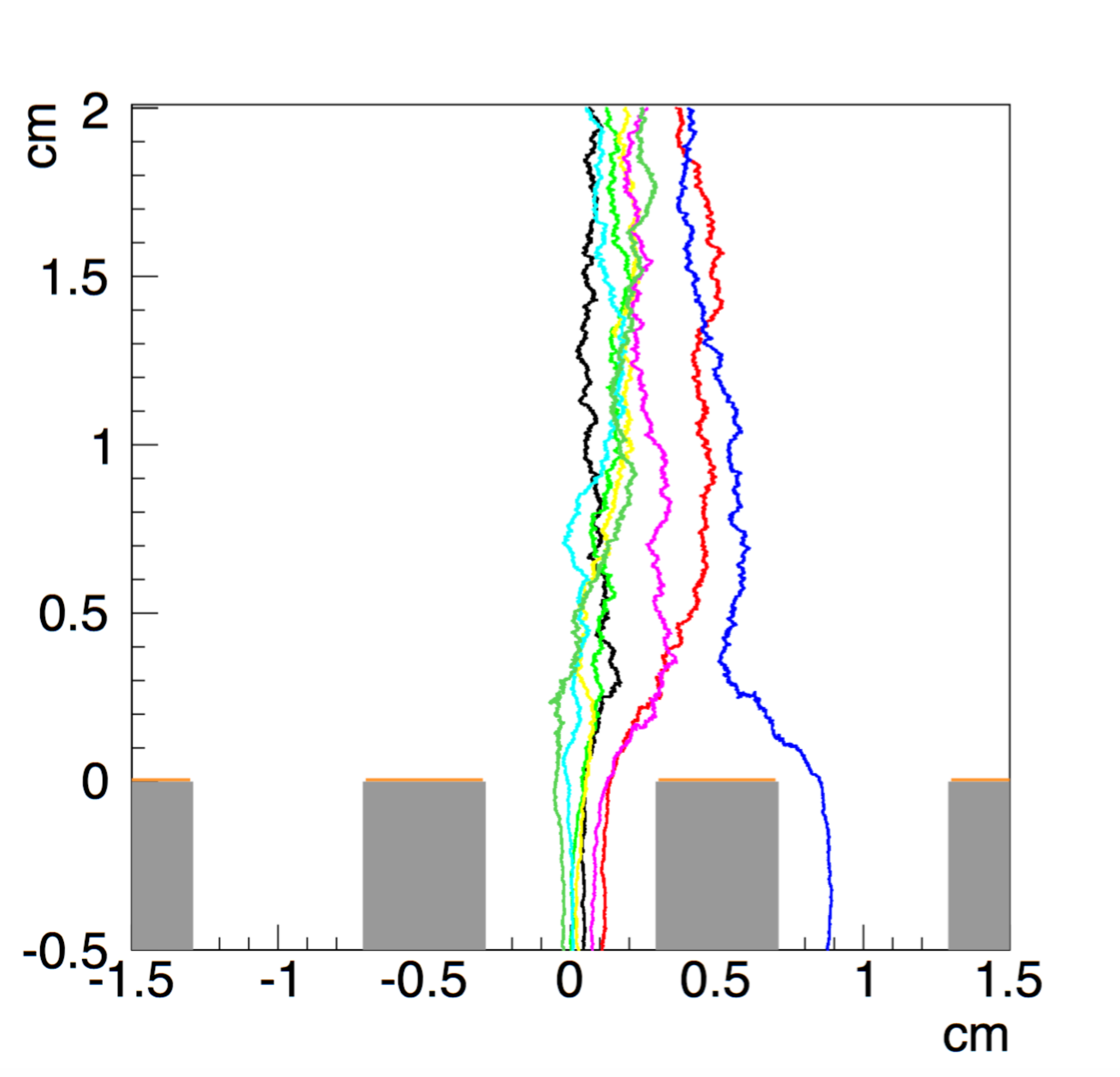} 
\caption{ Garfield++ simulation of path of drifting electrons to the ELCC.
Gray regions correspond to PTFE insulator. }
\label{fig:track}
\end{figure}

Due to diffusion effects while drifting, electrons do not always move along single electric
 field line.
Figure \ref{fig:track} shows examples of path of drifting electrons simulated with
 Garfield$++$\cite{ref:garfield}.
Among $1000$ tracks simulated above target cell, diffusion resulted in about $17\%$ entering a
nearest-neighbor cell and $0.6\%$ entered the nearest-diagonal cell.
The remainders are collected in the target cell.
Although the final positions of the electrons at the hole are blurred by diffusion, the energy
 resolution is not affected, since the EL light yield has little position dependence.
\par
The statistical fluctuation of the detected EL photon, as a fraction, is given as
\begin{equation}\label{eq:stat}
\sigma_{\rm total} = \sqrt{\sigma_{e^-}^2+\sigma_{\rm EL}^2} 
= \sqrt{\frac{W}{Q}\left( F+\frac{1}{g}\right)},\\
\end{equation}
 where, $\sigma_{e^-} = \sqrt{FW/Q}$ is fluctuation of number of ionized electrons and
 $\sigma_{\rm EL} = \sqrt{W/gQ}$ is fluctuation of number of EL photons,.
$W=22.1\,\rm eV$ and
 $F=0.14$ are $W-$value and fano factor of gaseous xenon, respectively, $Q=2458\,\rm keV$ is Q
 value of $0\nu\beta\beta$ decay of $^{136}\rm Xe$.
The $g$ is EL gain defined as the number of EL photons detected.
It is estimated that about $60$ photons reach $3\,\rm{mm} \times 3\,\rm{mm}$ SiPM for the ELCC
 with the optimized parameters by a Mopnte Carlo simulation.
Assuming mesh aperture ratio $R_{\rm mesh}=0.78$ and photon detection efficiency of SiPM
 $PDE=0.3$, the EL gain is calculated as $g=14$.
Since the gain required to achieve 0.5\%(FWHM) energy resolution is $2.8$ 
 from Equation \ref{eq:stat}, this EL gain is sufficiently large.

%% file: AXEL2016_Prototype.tex
\section{Prototype detector}
\label{sec:prototype}
\ We have produced a prototype detector with a 9 cm-long and 10 cm-diameter sensitive volume as shown in Fig.~\ref{fig:prototype}. The purpose of the prototype detector is to demonstrate the performance of the ELCC concept by measuring its energy resolution of the 511 keV gamma-ray's from a $^{22}$Na source. 
\subsection{Detection region}
\begin{figure}[htbp] \centering \includegraphics[width=60mm]{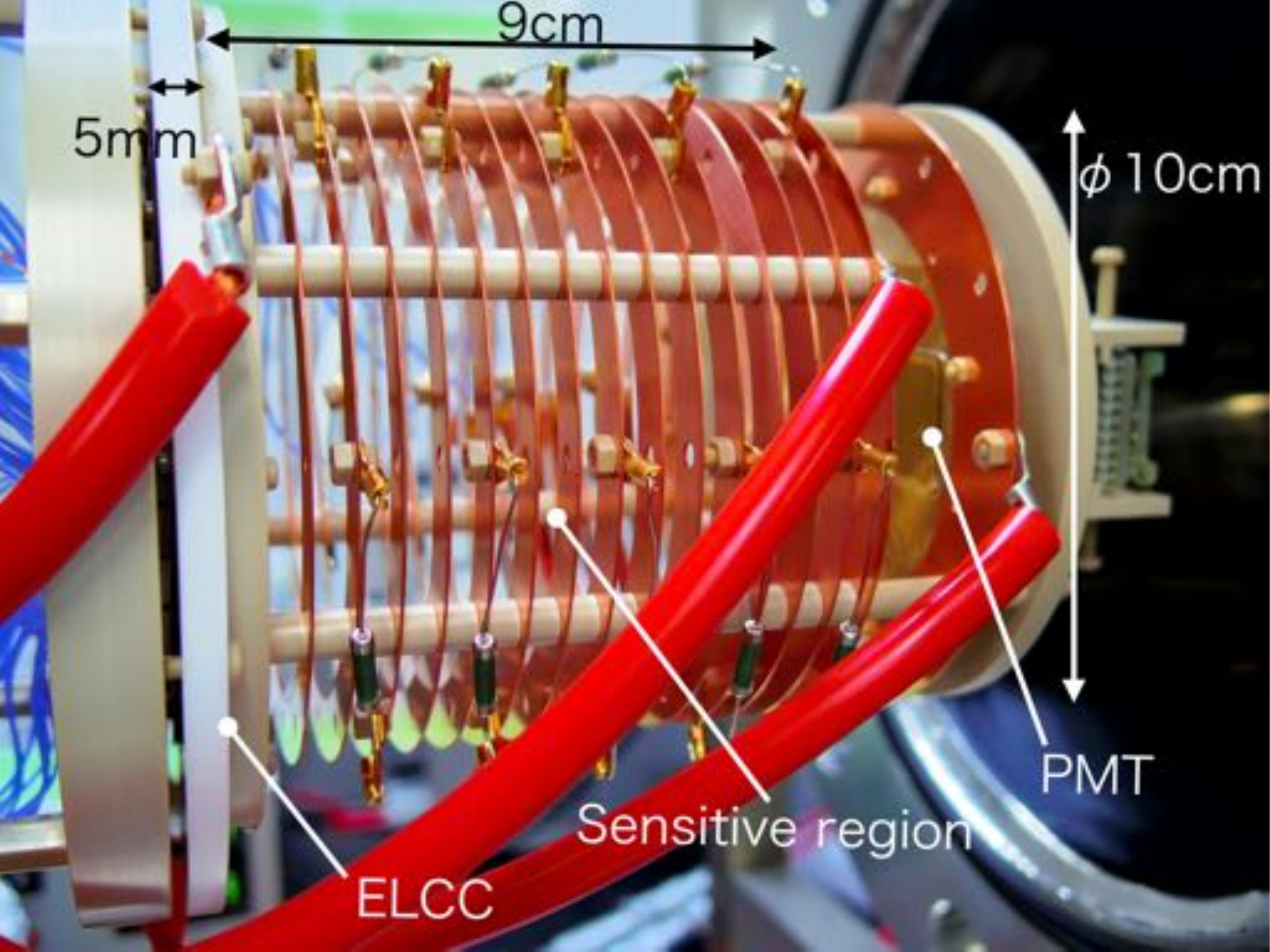} \caption{Picture of the prototype detector. The sensitive region, ELCC, PMT can be seen.}\label{fig:prototype}\end{figure}

\begin{figure}[htbp] \centering \includegraphics[width=70mm]{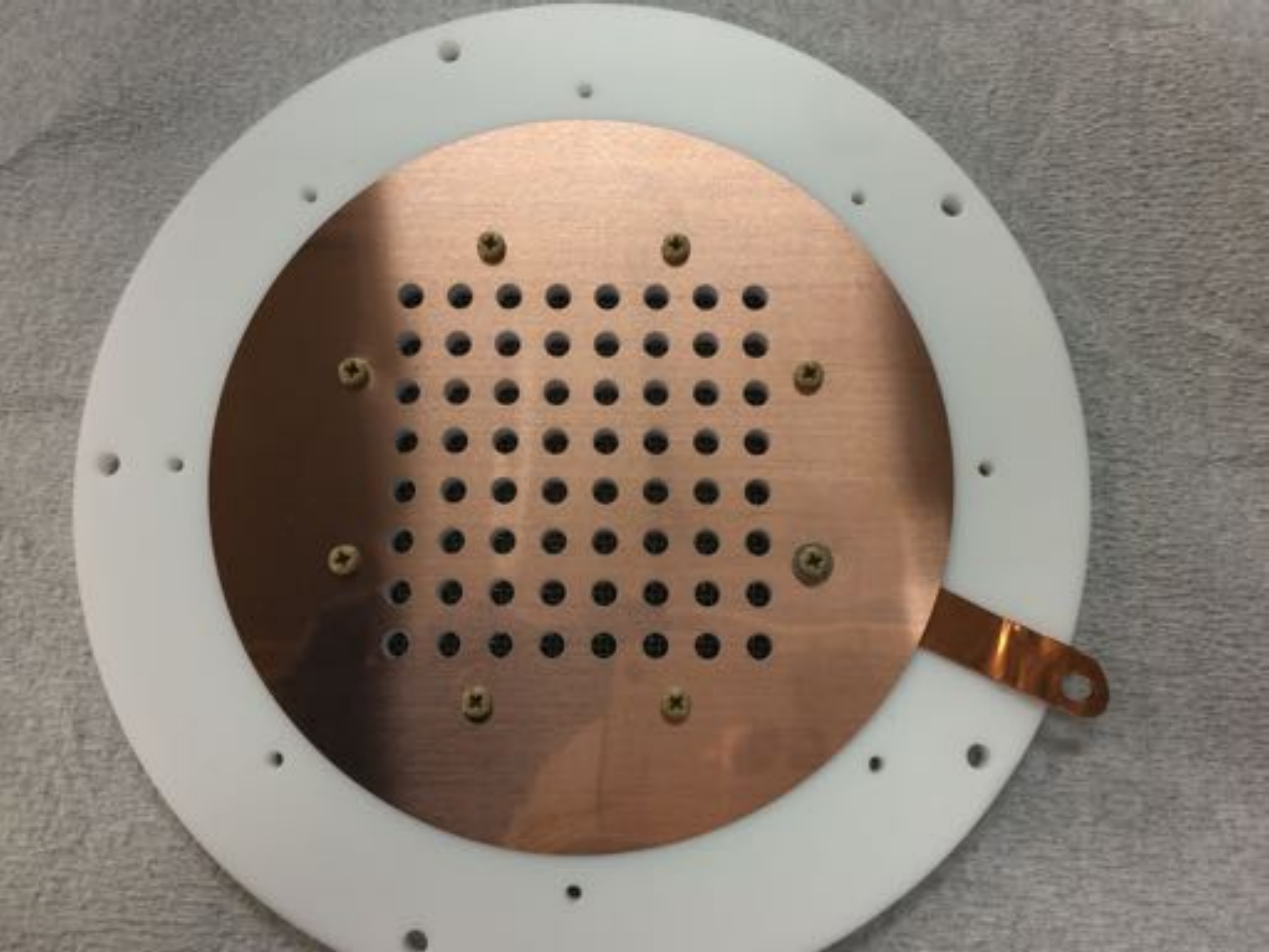} \caption{Picture of the ELCC of the prototype detector. The anode plate and PTFE body with 7.5 mm-pitch holes can be seen.}\label{fig:prototype_ELCC}\end{figure}

\ The detector has 64 ($8\times 8$) cells spaced with 7.5 mm pitch. Figure \ref{fig:prototype_ELCC} shows a picture of the ELCC of the prototype detector. The anode is made of a 0.1mm-thick oxygen-free copper plate with 4.0 mm-diameter holes. The ground mesh is made of a 0.3mm-diameter gilded tungsten wires with 100 mesh. The PTFE body is 5 mm-thick and has 3.8 mm-diameter holes (see Fig.~\ref{fig:prototype_ELCC}). An array of VUV-sensitive MPPCs (Hamamatsu Photonics S13370-4870) is attached behind the ground mesh, with an MPPC aligned with each hole. Each MPPC has a 3$\times$3 mm$^2$ sensitive area. Two PMTs (Hamamatsu photonics R8520-406MOD), which are sensitive to VUV photons and have a maximum pressure tolerance of 10 bar, are installed at the end of the detection region opposite that of the ELCC plane. Shaper rings which consist of fifteen 0.5mm-thick oxygen-free copper rings spaced at 5 mm intervals and connected to 100 M$\Omega$ resisters in series are used to create a uniform electric field in the drift region. These rings are spaced longitudinally along the axis of the drift region and the end which is close to the ELCC plane is connected to the anode plate through a 100 M$\Omega$ register. At the opposite side of the ends of the shaper rings, a mesh is spanned to create a uniform electric field and referred to cathode (see Fig.~\ref{fig:prototype}).\\
\ The electric field in the drift region is generated by applying a high voltage between the cathode and the anode plate. The applied electric field strengths are 100 V/cm/bar for the drift region and 2700 V/cm/bar for the EL region (between the anode plate and the grand mesh). 

\subsection{Pressure vessel and gas supply}
\ All the components of the prototype detector are housed in a pressure vessel made of stainless steel (SUS304). The vessel is designed to tolerate high pressure gas up to 10 bar. The inner diameter of the cylindrical section of the vessel is 208.3 mm, is 4mm thick, and is 340 mm long. It has two half-inch nipples with VCR joints for gas circulation. Both ends of the vessel are closed by JIS flanges. One of the flanges has feedthroughs with nine kapton sealed 25-bundles-ribbon-cables and 5 silicon-sleeved cables that withstand high voltage up to 30 kV. The ribbon cables are used to supply the bias voltages for the MPPCs and to read out the MPPCs' and PMTs' signals. The silicon-sleeved cables are used  to apply high voltage to cathode, anode and PMTs.\\ 
\ Xenon gas is filled into the vessel after passing through a molecular sieve filter and a getter filter for purification.



\subsection{Electronics and DAQ}
\begin{figure}[htbp] \centering \includegraphics[width=70mm]{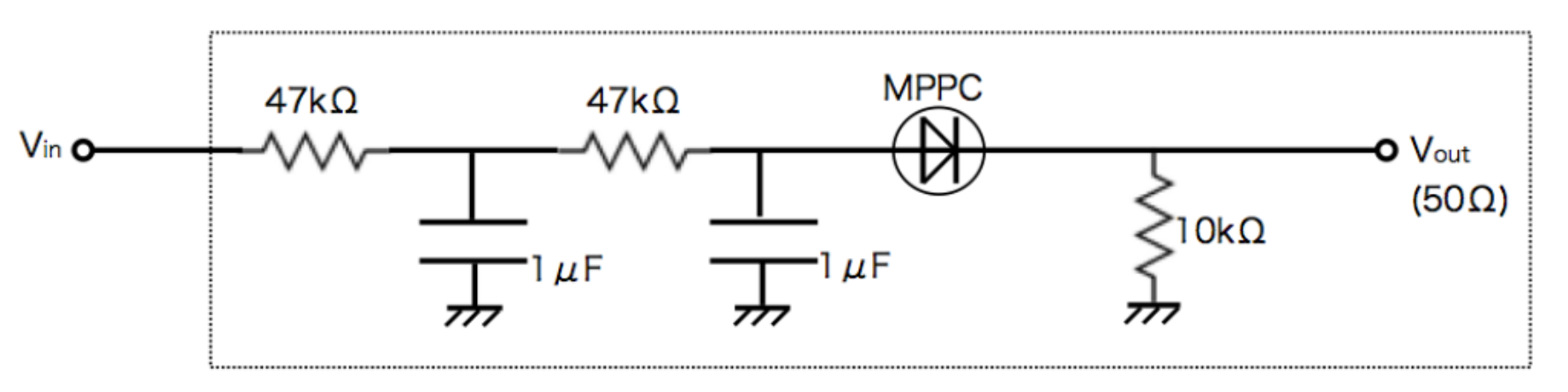} \caption{Circuit diagram of the low pass filter inserted between the bias power supply and a MPPC}\label{fig:LPF}\end{figure}

\ MPPC's having same break down voltage within $\pm$ 0.8 V were selected at the delivery from the manufacturer. Accordingly, the same bias voltage is applied to 64 channels with a single DC power supply. To suppress potential noise from the power supply and to prevent crosstalk among channels, each MPPC is equipped with a low pass filter (LPF) on the bias line, as shown in Fig.~\ref{fig:LPF}. 
The time constant and capacitance of the LPF have been adjusted to 15 msec and 1 $\mu$F to produce wide signal pulses and large charges, typically a few $\mu$sec and up to $\sim$10$^5$ photons per channel.

\begin{figure*}[htbp] \centering \includegraphics[width=140mm]{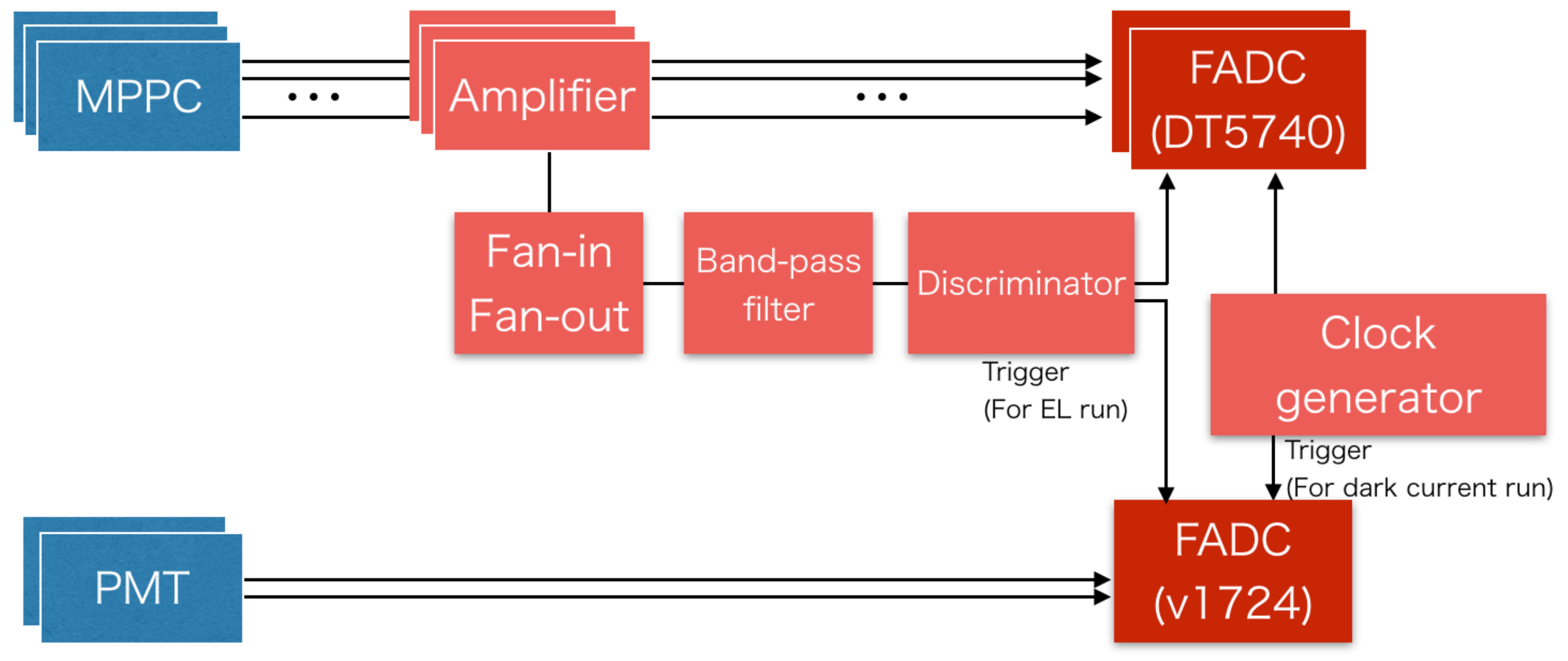} \caption{Data acquisition diagram.}\label{fig:DAQlogic}\end{figure*}

\ A schematic diagram of the data acquisition system (DAQ) is shown in Fig.~\ref{fig:DAQlogic}. Signals from MPPCs and PMTs are recorded by waveform digitizer modules. MPPC signals are amplified by a factor of 10 before being recorded with two 32ch-12bit-65MHz-sampling digitizer modules (DT5740 by CAEN inc.). One 8ch-14bit-100MHz-sampling digitizer module (v1724 by CAEN inc.) is used for signals from the PMTs. Waveform data are recorded at 6000 samples (96 $\mu$sec) for MPPC signals and 10000 samples (100 $\mu$sec) for PMT signals. The three modules are linked optically and controlled by a PC. The sum of fiducial MPPC signals shown in Fig.\ref{fig:MPPCconf} is formed by linear fan-in fan-out modules using a secondary outputs on the the amplifiers and is then fed to a band-pass filter (BPF) and discriminated by a NIM module to create the DAQ trigger signal.
The BPF's frequency range is between $10^3$ and $10^6$ Hz, which eliminates dark current pulses from the MPPCs but allows acquisition of electroluminescence signals.\\
\ A pulse generator is also used to generate a trigger signal to take dark current data used to determine the gain of the MPPCs.

%% file: AXEL2016_Analysis.tex
\section{Analysis of the prototype detector data} 
\label{sec:analysis}
\ \ We evaluated the energy resolution of the prototype detector at 4 bar using the 122 keV gamma-ray from a $^{57}$Co source. 
\subsection{MPPC gain calibration}
\begin{figure}[htbp]  
 \centering \includegraphics[width=60mm]{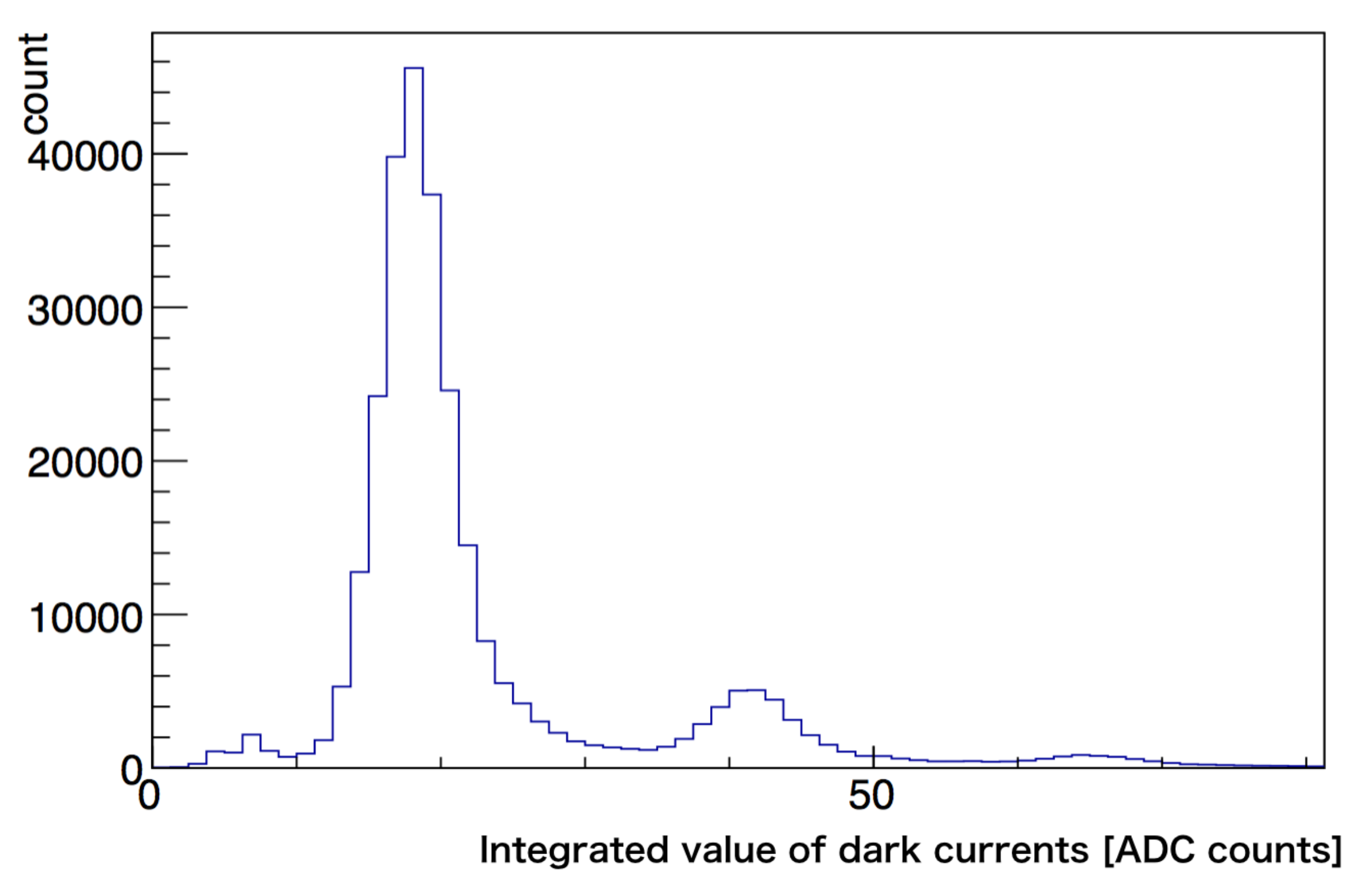} \caption{Typical distribution of the charge of dark currents.}\label{fig:opinteg}\end{figure}

\begin{figure}[htbp]
 \centering \includegraphics[width=65mm]{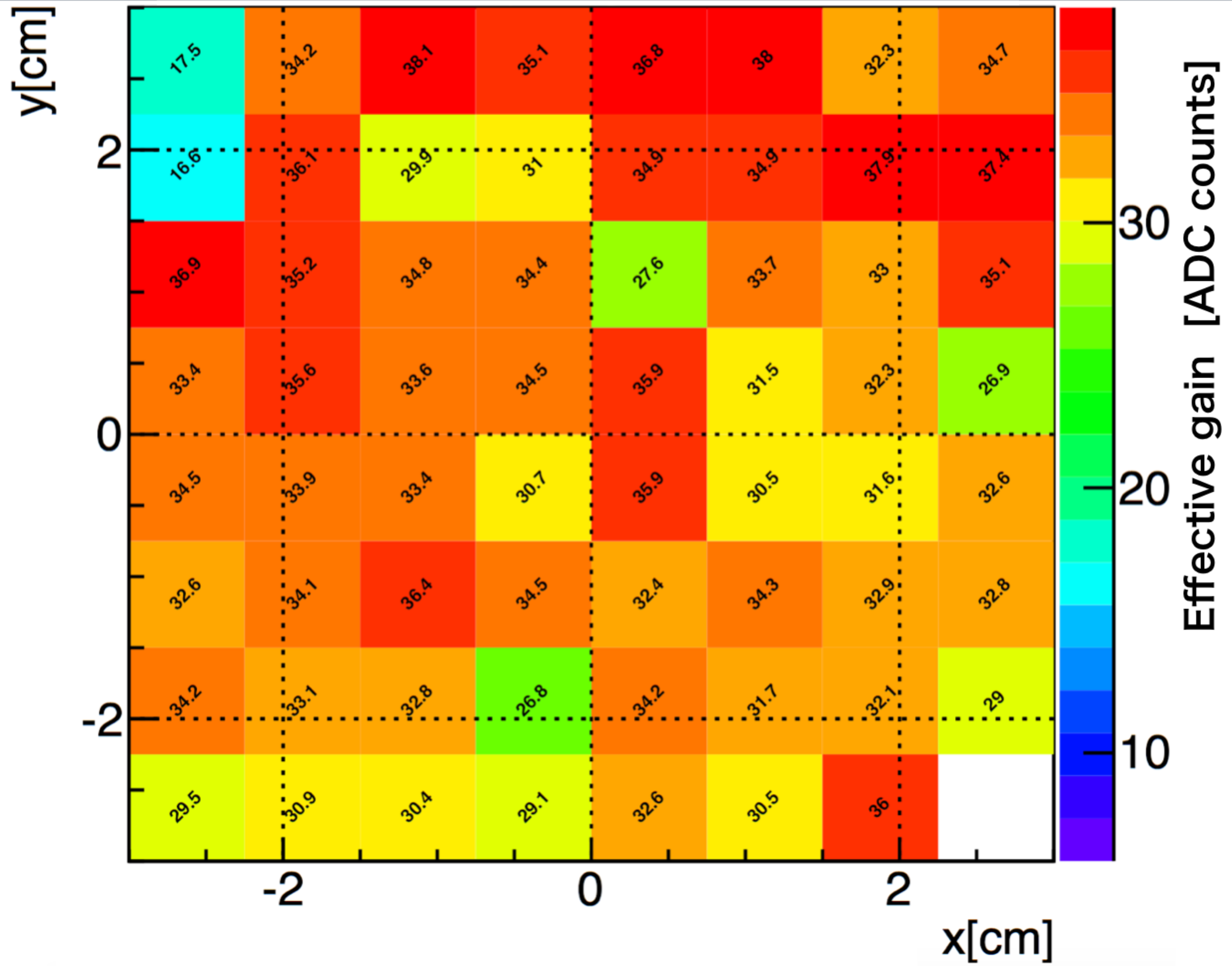} \caption{Gain map of MPPCs. The two top left channels and the bottom right channel are dead.}\label{fig:MPPCgain}\end{figure}

\ Each MPPC's gain is determined using its dark current. An example dark current charge distribution from a single MPPC is shown in Fig. \ref{fig:opinteg}. Peaks corresponding to one, two and three photo equivalent (p.e.)  are clearly seen. These peaks are fitted with Gaussians in order to determine a 0.5 p.e. threshold. The mean charge of the events above the threshold is taken as ``effective gain'', which corresponds to the average gain after crosstalk and afterpulse effects of MPPC are taken into account. The obtained gain map is shown in Fig. \ref{fig:MPPCgain}.  Using this effective gain, the integral of the signal from each channel is translated to the number of photons.\\
\ Dark current rate is also calculated by counting the number of dark current pulses. 

\subsection{Hit channel determination and integration}
\begin{figure}[htbp] \centering \includegraphics[width=70mm]{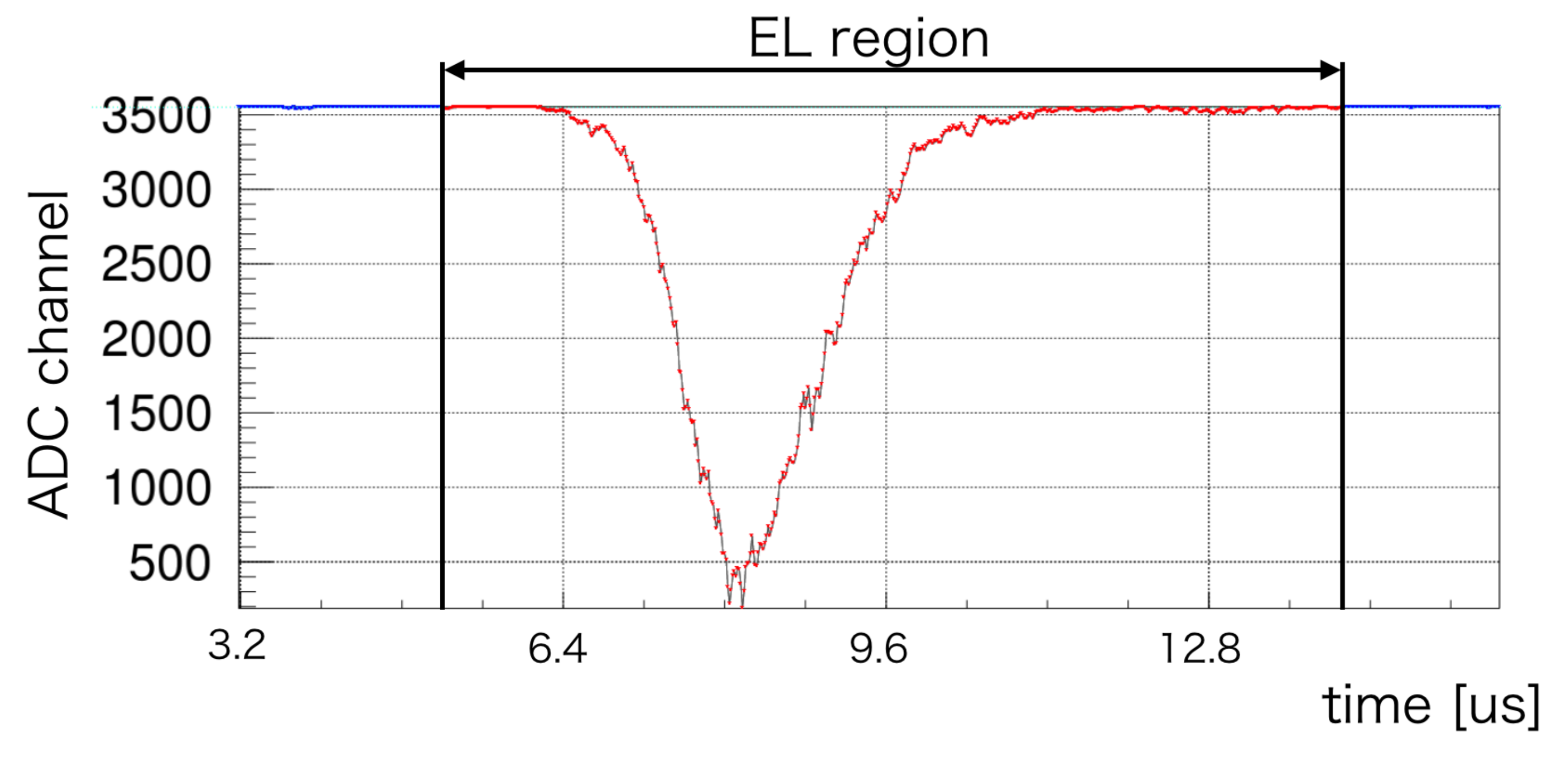} \caption{Example waveform with an electroluminescence light signal. The region between the two solid vertical lines represents the EL region.}\label{fig:waveform_ex}\end{figure}

\ For each event and for each channel, the baseline is determined as the truncated mean of FADC counts with only 4 counts around the peak to avoid being affected by the EL. 
The baseline's r.m.s ($\sigma$) is used to set an analysis threshold of 3$\sigma$ above the baseline for signal integration. \\
\ For each MPPC channel the total integrated FADC counts $S$[counts] is calculated by integrating the differences between the baseline and the FADC counts throughout all samples in one event. The channels which meet the condition
\begin{equation} S[\textrm{counts}] > \alpha \times Q_\textrm{dark} \end{equation}
are regraded as ``hit channel''.
$Q_{\textrm{dark}}$ is expected dark charge over all samples in an event and calculated from the dark current rate. The constant $\alpha$ is set to 1.1 in this analysis.\\

\ For each hit channel, the EL region is defined as follows : 
\begin{enumerate}
\item The waveform is smoothed by averaging over the 50 neighboring samples in order to avoid to select dark current signal of MPPCs. The maximum point of the resulting waveform is selected as the point to start searching for the signature of EL. The smoothed waveform is only used to select this starting point.
\item Starting from the point selected in step 1, the points where the waveform falls below the analysis threshold for at least 40 continuos samples is searched toward both sides. 
\end{enumerate}
Figure \ref{fig:waveform_ex} shows an example waveform with its EL region.
The integrated number of counts in the EL region after subtracting the baseline is converted to  photon counts by dividing by the gain of the MPPCs.
The total number of photons in an event is obtained by adding the photon counts of all hit channels.

\subsection{Cuts and corrections}
\begin{figure}[htbp] \centering \includegraphics[width=70mm]{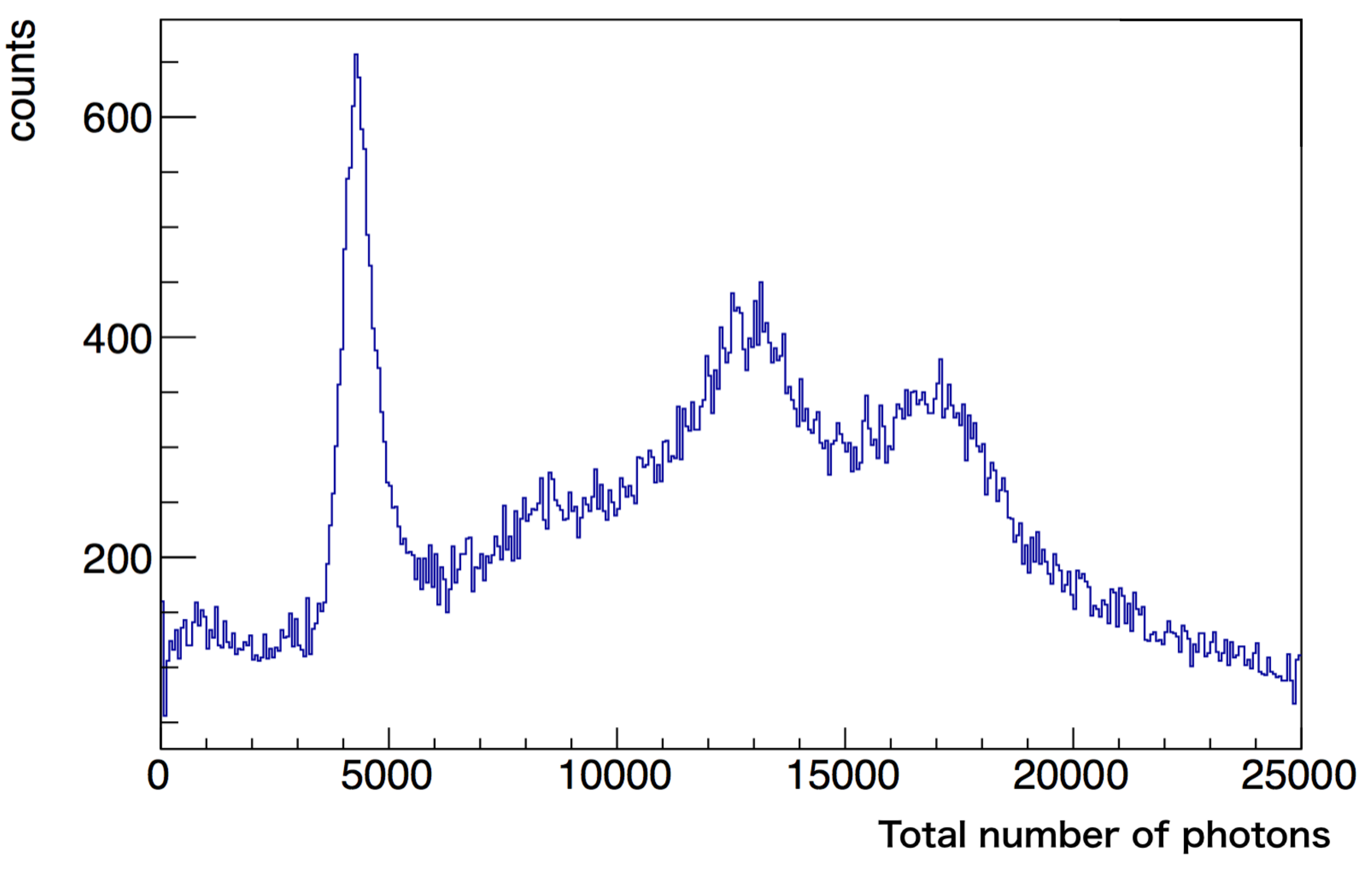} \caption{Photon count distribution before any cuts or correction are applied.}\label{fig:RawPhoton}\end{figure}

\ Figure \ref{fig:RawPhoton} shows the obtained photon count distribution without any cuts or corrections. 

\begin{figure}[htbp] \centering \includegraphics[width=45mm]{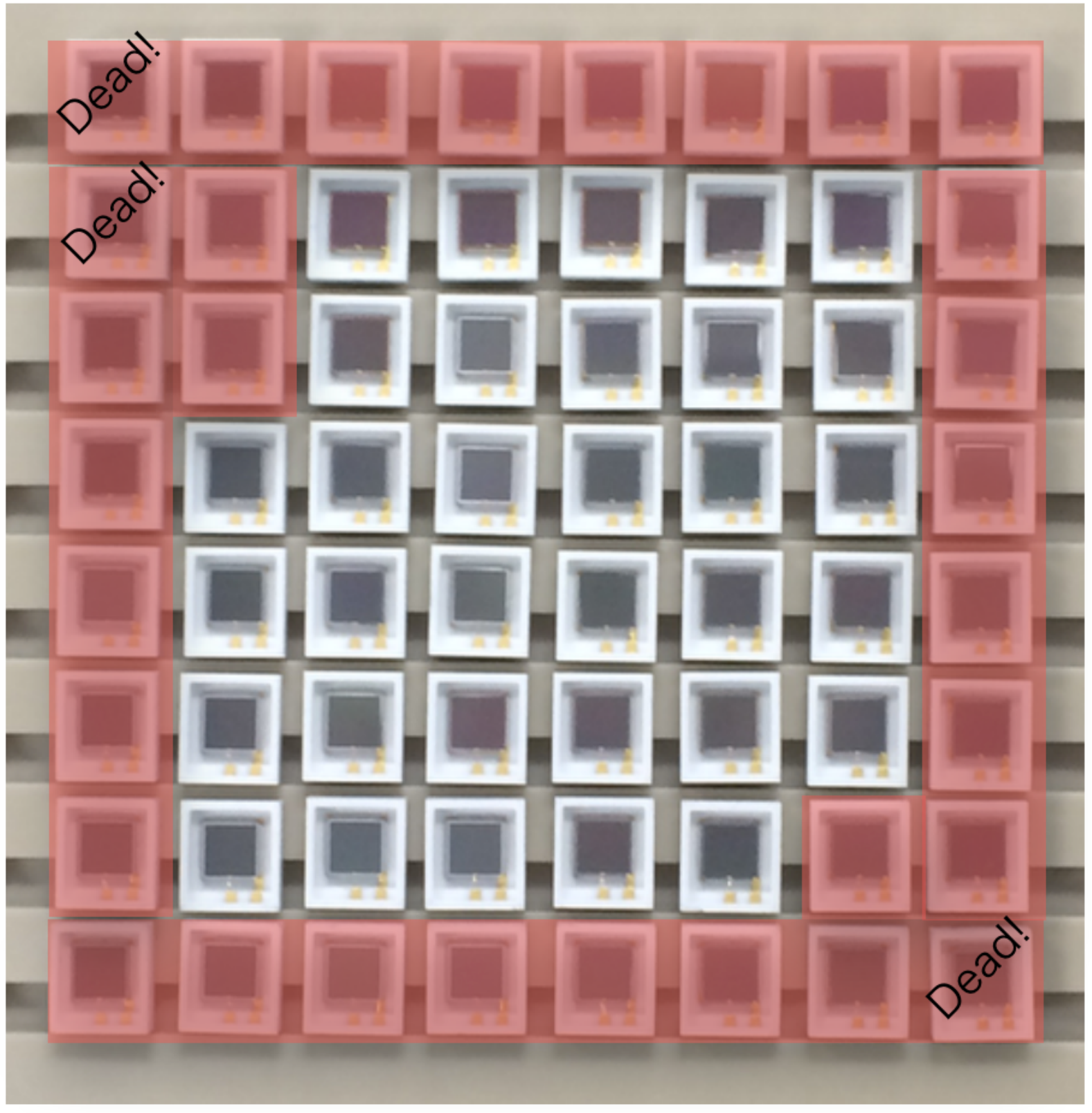} \caption{Configuration of MPPCs. Red region is used for veto signal.}\label{fig:MPPCconf}\end{figure}

\begin{figure}[htbp] \centering \includegraphics[width=80mm]{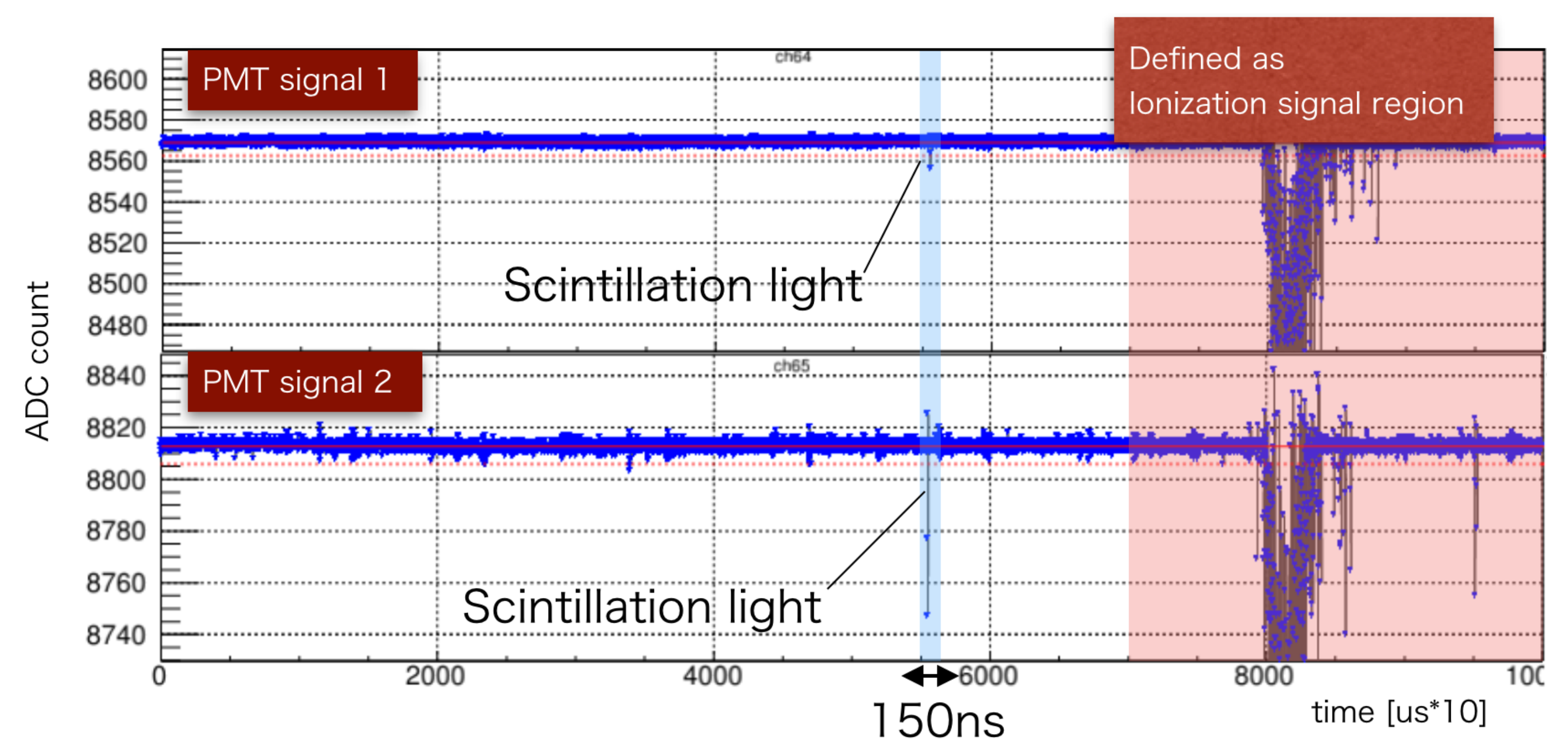} \caption{Example of PMT waveforms satisfying the coincidence condition.}\label{fig:PMTcoin}\end{figure}

Several cuts and corrections are applied to this distribution. Events contained in the center 33 cells are selected (see Fig.\ref{fig:MPPCconf}). Events with saturated ADC values are removed. Coincidence of two PMT signals within 150 nsec is required in order to distinguish scintillation light signal from dark current noise of PMTs. Typical PMT waveforms satisfying the coincidence condition are shown in Fig.\ref{fig:PMTcoin}. Using the time difference between the timing of coincidence signal of two PMTs and the timing of the MPPC signals, the event position along the drift direction (z axis) is reconstructed and a fiducial cut along z axis is applied. Events contained in 2 cm - 7.5 cm region along z direction away from the anode plate are selected.

\begin{figure}[htbp] \centering \includegraphics[width=65mm]{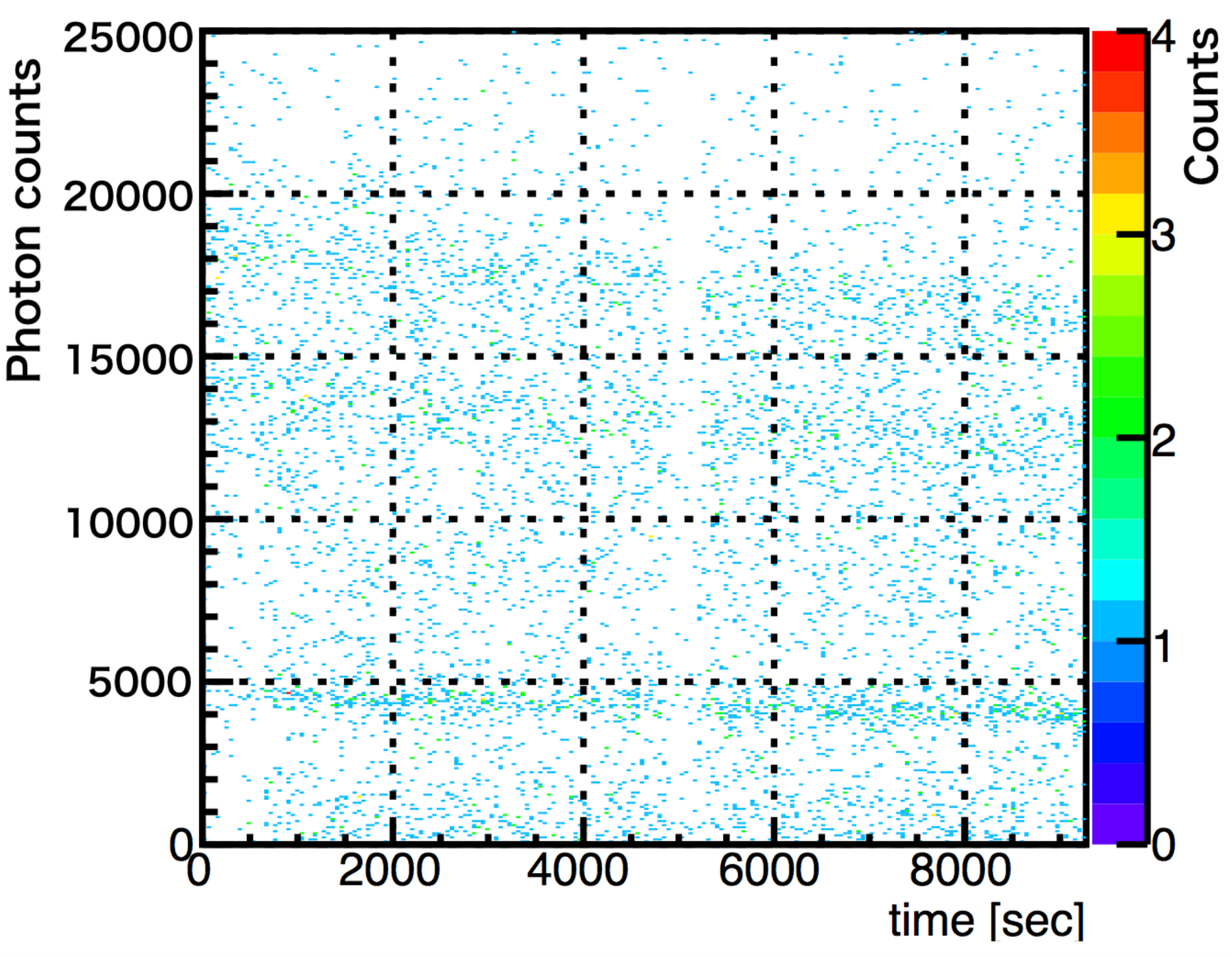} \caption{Dependence of the light yield on elapsed time.}\label{fig:time_vs_photon}\end{figure}

Figure \ref{fig:time_vs_photon} shows the observed number of photons as a function of time. The light yield decreased as time elapsed. This is considered to be caused by increasing impurities in xenon gas. A correction is applied to compensate for this decrease.

\begin{figure}[htbp] \centering \includegraphics[width=55mm]{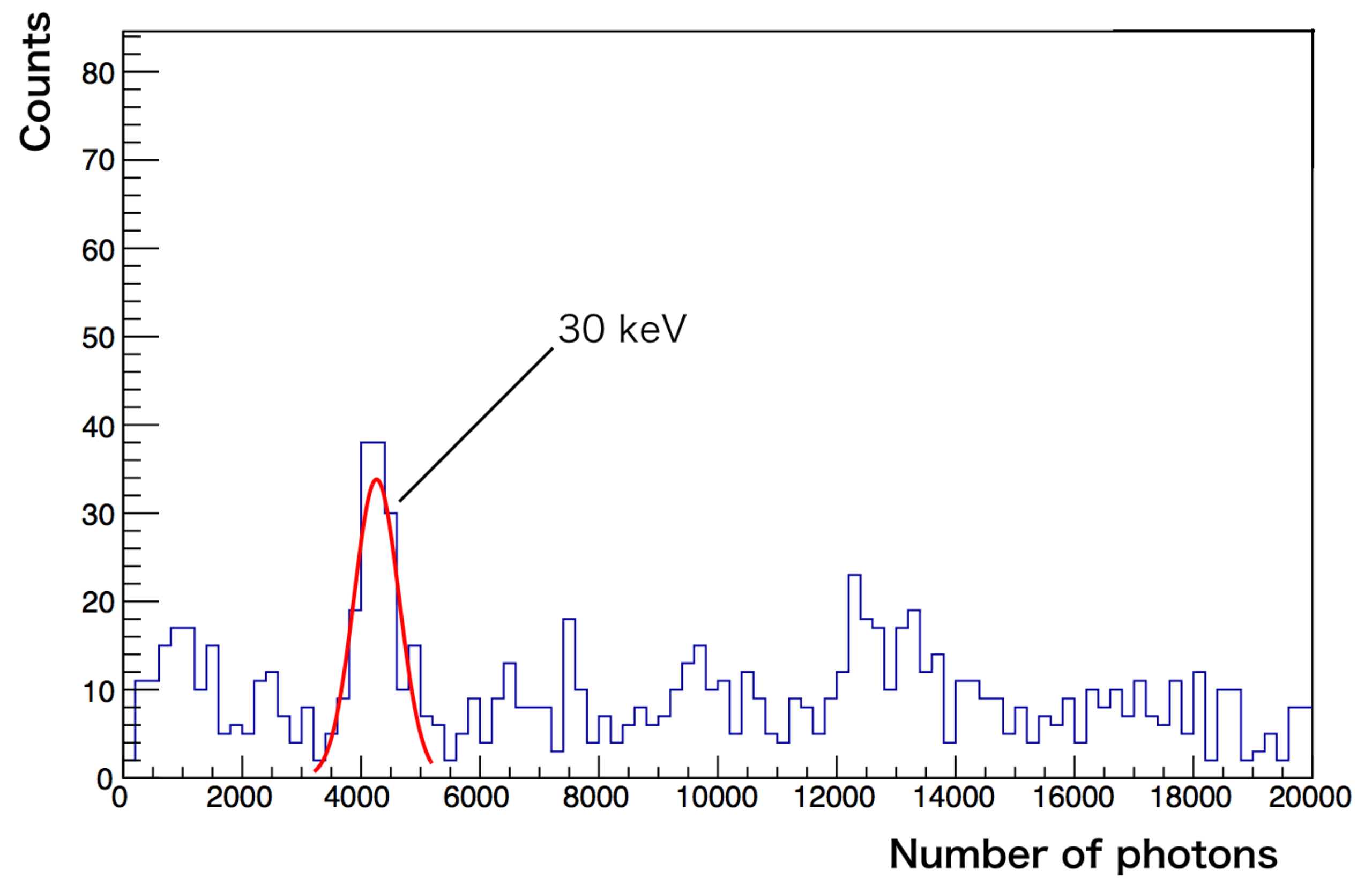} \caption{Example of the cell gain calibration distribution. Red curve represents the Gaussian fit result. }\label{fig:ELgaincorrection_method} \end{figure}

Calibration of the EL gain of each channel is done cell by cell. For each cell events are selected in which it had the highest number of observed photons and in which no cells other than its four nearest neighbors were hit.  
The distribution obtained by summing the number of photons detected by these cells shows clear 30 keV characteristic X-ray peak from Xenon as shown in Fig.\ref{fig:ELgaincorrection_method}. The gain of each cell is determined using this peak.

\ Finally, the contribution of the MPPC dark current to the electroluminescence signal is subtracted channel-by-channel.

%% file: AXEL2016_Results.tex
\section{Performance of the prototype detector} 
\label{sec:results}
\subsection{Energy resolution}
\begin{figure}[htbp]  
\centering \includegraphics[width=75mm]{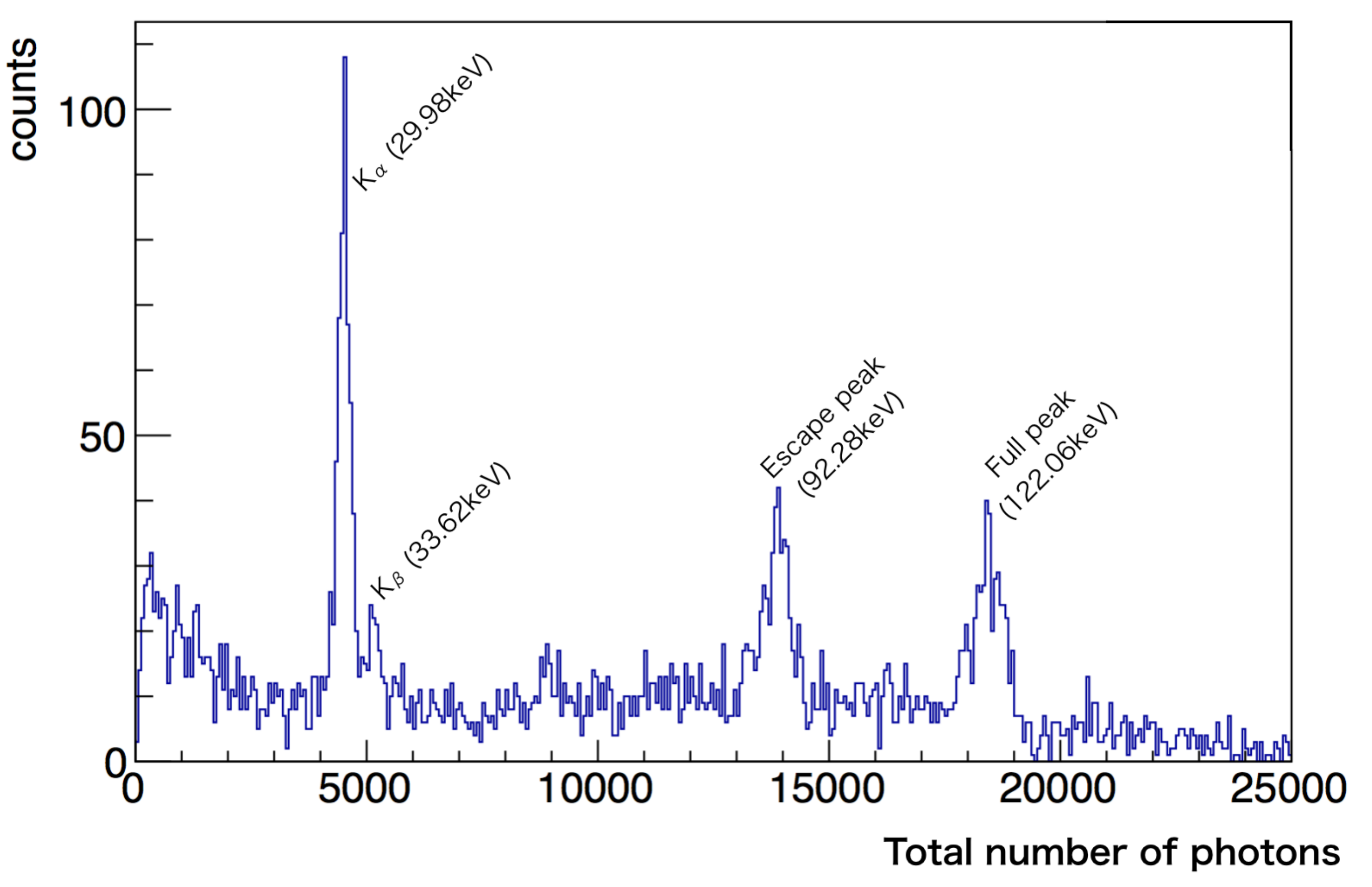} \caption{Number of detected photons spectrum when irradiated with a 122 keV gamma-rays from $^{57}$Co source after all cuts and corrections.}\label{fig:photon_fit}\end{figure}

\begin{figure}[htbp]   \centering \includegraphics[width=75mm]{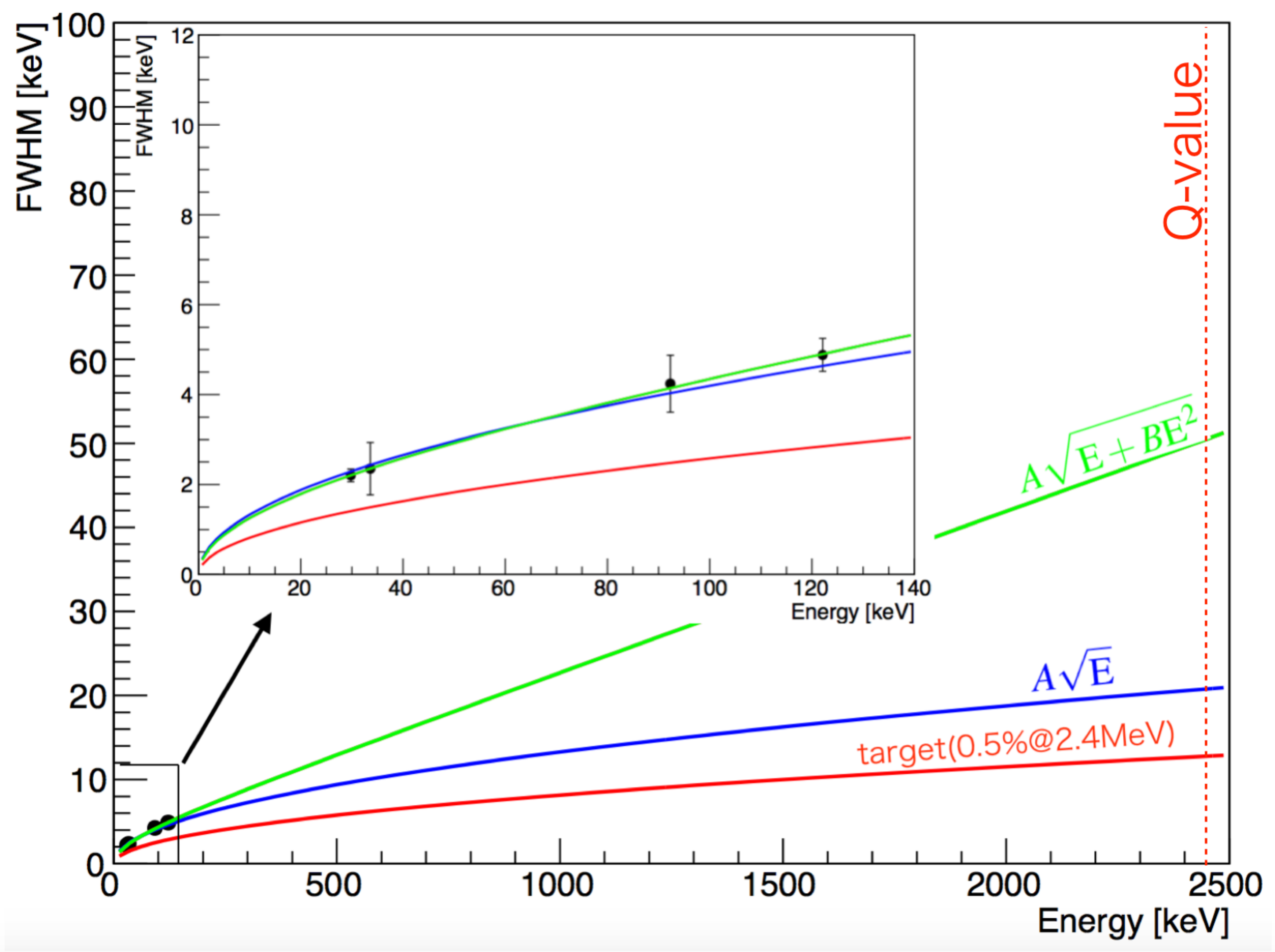} \caption{Energy resolution as a function of deposited energy. Lines show the results of fits to the data using the functions $A\sqrt{E}$ and $A\sqrt{E+BE^2}$ described in the text.}\label{fig:6} 
\end{figure}

\ Figure \ref{fig:photon_fit} shows the photon count distribution for the $^{57}$Co source with all cuts and corrections applied. To evaluate the energy resolution, the first three peaks (29.8keV, 33.0keV, 92keV) are fitted with Gaussians and the last peak (122keV) with ``Gaussian $+$ linear function'' to account for contributions from the background. The obtained energy resolution is summarized in Table \ref{Table1}. To estimate the energy resolution at the $0\nu\beta\beta$ Q-value of Xenon, the measured resolutions at the four peaks are fit with under two energy-dependence assumptions and extrapolated to 2458 keV. The first assumes the resolution depends only on the statistical uncertainty, $A\sqrt{E}$, and the second assumes an additional linear dependence, $A\sqrt{E+BE^2}$, where $E$ is the deposited energy in keV, and $A$ and $B$ are fitting parameters.
The fit results are  $(0.42\pm 0.019)\sqrt{E}$ and  $(0.39\pm 0.036)\sqrt{E+(0.0023\pm 0.0028)E^2}$ respectively and are shown in Fig.\ref{fig:6}. The extrapolated energy resolution (FWHM) at 2458 keV is 0.85$\%$ with the function A$\sqrt{E}$ and 2.03$\%$ with the function $A\sqrt{E+BE^2}$.\\
\ The 0.85\% energy resolution is comparable to those obtained in \cite{ref:NEXT_511keV} and \cite{ref:micromegas}. In those measurements, the energy resolutions extrapolated to 2458 keV by assuming $A\sqrt{E}$ are 0.83\% and 0.9\%, respectively. A drift chamber in the scintillation (electroluminescence) mode readout, filled with 0.9 MPa Xenon, showed good energy resolution in 1997\cite{ref:ScintiDC}. The achieved energy resolution is 2.7\% (FWHM) for 122 keV.
We are planning to try to increase gas pressure and increase electric field in order to achieve better energy resolution.
At the current EL gain (about 3.5 per ionization electron), there is non-negligible contribution from the EL conversion fluctuation to the energy resolution (equation(2)). The EL light gain is higher at higher pressure and the energy resolution would be improved at, for example, 10 bar. In this measurement, the required electric field at the ELCC to fully collect electrons is 3 kV/cm/bar, while actually applied field was limited to 2.7 kV/cm/bar due to discharge. We are modifying the ELCC structure so that higher voltage can be applied. In addition, we are planning to measure at the higher energy to make more accurate estimate the energy resolution at the $0\nu\beta\beta$ Q-value.

\begin{table}[htb]
\caption{Energy resolution of each peak from the $^{57}$Co source. Errors are statistical only.} 
\centering
\begin{tabular}{c||c|c} \hline
        Energy &Photon count & Resolution(FWHM) \\ \hline
        28.78 keV & 4517.3   & 7.3$\pm$ 0.47\% \\ 
        33.62 keV & 5169.5   & 7.0$\pm$ 1.7\%   \\
        92.28 keV & 13900.2 & 4.6$\pm$ 0.69\%   \\
        122.0 keV & 18445.0 & 4.0$\pm$ 0.30\%   \\
\end{tabular}
\label{Table1}
\end{table}

 